\documentclass[10pt,runningheads]{llncs}

\usepackage[british]{babel}
\usepackage[T1]{fontenc}
\usepackage[utf8x]{inputenc}

\usepackage{lmodern}

\usepackage{fix-cm}\rmfamily 
\DeclareFontShape{T1}{lmr}{b}{sc}{<->ssub*cmr/bx/sc}{}
\DeclareFontShape{T1}{lmr}{bx}{sc}{<->ssub*cmr/bx/sc}{}

\usepackage{amsmath,amsfonts,amssymb,amsthm}
\usepackage{subcaption,graphicx,wrapfig}
\usepackage{xspace,xfrac,xifthen}
\usepackage[dvipsnames,table]{xcolor}
\usepackage[inline]{enumitem}
\usepackage{booktabs,multirow,array}
\usepackage[normalem]{ulem}
\usepackage[symbol*,hang,marginal]{footmisc}  
\usepackage{tikz,tikzsymbols,pgfkeys,adjustbox}
\usepackage{relsize}   
\usepackage{ragged2e}  
\usepackage{breakcites}
\usepackage{csquotes}
\usepackage{marginnote,tokcycle,setspace}
\usepackage{listings}
\usepackage{graphicx}
\usepackage{tabularx}
\usepackage{lipsum}

\usepackage[normalem]{ulem}
\useunder{\uline}{\ul}{}
\usepackage{pdflscape,longtable}  

\RequirePackage[%
  bookmarks,
  unicode,
  colorlinks=true,
  allcolors=blue!65!black!90,
  breaklinks=true]{hyperref}
\usepackage[nameinlink]{cleveref}

\graphicspath{{figures/},{results/},}  

\newboolean{tosubmit}
\setboolean{tosubmit}{true}

\newboolean{showlinums}
\setboolean{showlinums}{false}
\ifthenelse{\boolean{showlinums}}{%
  \usepackage[pagewise]{lineno}  
  \linenumbers
  
}{}

\ifthenelse{\boolean{tosubmit}}{}{ 
  \usepackage{draftwatermark}  
  \SetWatermarkColor{Blue!11}
}

\Crefname{equation}{eq.}{eqs.}
\crefname{equation}{equation}{equations}
\Crefname{figure}{Fig.}{Figs.}
\crefname{figure}{figure}{figures}
\Crefname{tabular}{Table}{Tables}
\crefname{tabular}{table}{tables}
\Crefname{definition}{Def.}{Defs.}
\crefname{definition}{definition}{definitions}
\Crefname{proposition}{Prop.}{Props.}
\crefname{proposition}{proposition}{propositions}
\Crefname{section}{Sec.}{Secs.}
\crefname{section}{section}{sections}
\crefname{listing}{code}{code\ blocks}

\setquotestyle{latin}
\let\oldmktextquote\mktextquote
\renewcommand*{\mktextquote}[6]{\oldmktextquote{#1}{\textit{#2}}{#3}{#4}{#5}{#6}}

\DefineFNsymbols*{lamport}[math]{
  {\hspace{2pt}\reflectbox{\bfseries\textcopyright}\hspace{1pt}}
  {\hspace{1pt}\dagger\hspace{.2pt}}
  {\hspace{1pt}\ddagger\hspace{.2pt}}
  {\hspace{1pt}\ast\hspace{.2pt}}
  {\hspace{1pt}\S\hspace{.2pt}}
  {\hspace{1pt}\P\hspace{.2pt}}
  {\hspace{1pt}\|\hspace{.2pt}}
  {\hspace{1pt}\dagger\dagger\hspace{.2pt}}
  {\hspace{1pt}\ddagger\ddagger\hspace{.2pt}}
  {\hspace{1pt}\ast\ast\hspace{.2pt}}
}
\setfnsymbol{lamport}

\lstdefinestyle{solidity}
{
    language={Solidity},
    basicstyle=\scriptsize\ttfamily,
	columns=fullflexible,keepspaces,
	backgroundcolor=,  
    numbers=left,
    numbersep=1em,
    numberstyle=\tiny\color{black!75},
	numberblanklines=false,
    xleftmargin=4\dimexpr\fboxsep+\fboxrule\relax,
    xrightmargin=\dimexpr\fboxsep+\fboxrule\relax,
    breaklines=true,
    tabsize=2,
}

\makeatletter
\renewcommand{\paragraph}{\@startsection{paragraph}{5}{0em}%
  {.7ex plus .2ex minus .1ex}%
  {-.5em}%
  {\bfseries}}
\makeatother

\makeatletter
   \def\@citecolor{blue}%
   \def\@urlcolor{blue}%
   \def\@linkcolor{blue}%

\def\orcidID#1{\smash{\href{http://orcid.org/#1}{\protect\raisebox{-.1ex}%
{\protect\includegraphics[width=8pt]{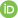}}}}}
\makeatother

\makeatletter
\def\THICKhrulefill{\leavevmode \leaders \hrule height 5pt\hfill \kern \z@}
\def\getfirst#1#2\relax{\tctestifnum{\count@stringtoks{#1}>1}{ERROR}{#1}}
\makeatother
\newcommand{\colorpar}[3]{\colorbox{#1}{\parbox{#2}{#3}}}
\newcommand{\marginremark}[3]{%
  \ifthenelse{\boolean{tosubmit}}{}{
	\marginnote{\raggedrightmarginnote\colorpar{#2}{.8\linewidth}%
      {\raggedrightmarginnote\color{#1}#3}}
}}
\newcommand{\textremark}[5]{%
  \ifthenelse{\boolean{tosubmit}}{}{
  \marginremark{#1}{#2}{\tiny\sffamily{[#3]\ #5}}%
  {\def\ULthickness{.8pt}\color{#1!80!black}\uline{#4}}
}}
\newcommand{\highlightedremark}[4]{%
  \ifthenelse{\boolean{tosubmit}}{}{
	\begin{center}\fcolorbox{#1}{#2}{%
	\begin{minipage}{.98\linewidth}\color{#1}%
	\textbf{\THICKhrulefill[ #3 ]\THICKhrulefill}%
	\par\noindent#4\end{minipage}}\end{center}%
}}
\newcommand{\hey}[4]{%
  \ifthenelse{\boolean{tosubmit}}{}{
  \reversemarginpar
  \leavevmode\marginnote{\sffamily\Large\color{#1}@\getfirst#3\relax\relax}
  \colorbox{#2}{\sffamily\bfseries{@#3:}}~{\sffamily\color{#1}#4}}
  \normalmarginpar}
\newcommand{\todo}[1]{%
  \ifthenelse{\boolean{tosubmit}}{}{
  \noindent\textsf{\color{Red!93!black}\textbf{TODO:} #1}%
  \marginnote{\textsf{\color{red}\bfseries TODO}}}}
\newcommand{\tocite}[1][??]{%
  \ifthenelse{\boolean{tosubmit}}{}{
  \noindent\textbf{\sffamily\textcolor{blue!85}{[#1]}}%
  \marginnote{\textsf{\color{blue}\bfseries CITE!}}}}
\colorlet{TO-fg}{BrickRed}
\colorlet{TO-bg}{orange!11}
\colorlet{CEB-fg}{TealBlue!75!green!75!black}
\colorlet{CEB-bg}{Aquamarine!8}

\crefname{example}{example}{examples}

\usetikzlibrary{positioning,tikzmark,fit,patterns,patterns.meta}

\colorlet{hlboxcoldraw}{black!65}  
\colorlet{hlboxcolfill}{black!6}   
\newsavebox\BODYBOX  
\def\BODY{\unhbox\BODYBOX}
\NewDocumentEnvironment{hlbox}{ O{} }
  {\vspace{1ex plus .5ex minus .3ex}%
	\begingroup\centering\begin{spacing}{1.05}%
	\begin{lrbox}{\BODYBOX}\begin{minipage}{.96\linewidth}%
	\ifthenelse{\equal{#1}{}}{}{\textbf{#1}}}
  {\end{minipage}\end{lrbox}\begin{tikzpicture}%
	\node[rectangle,rounded corners=.3mm,inner sep=1.3ex,thick,
	      draw=hlboxcoldraw,fill=hlboxcolfill] {\BODY};
	\end{tikzpicture}\end{spacing}\endgroup%
	\vspace{1ex plus .5ex minus .3ex}}

\newcolumntype{L}[1]{>{\RaggedRight\let\newline\\\arraybackslash}m{#1}}
\newcolumntype{P}[1]{>{\RaggedRight\let\newline\\\arraybackslash}p{#1}}
\newcolumntype{C}[1]{>{\Centering\let\newline\\\arraybackslash}m{#1}}
\newcolumntype{R}[1]{>{\RaggedLeft\let\newline\\\arraybackslash}m{#1}}

\colorlet{shade1}{black!11}
\colorlet{shade2}{white}
\colorlet{shade3}{Black!11}  
\colorlet{shade4}{Black!6}  

\def\colcolortablepreamble{%
  \renewcommand{\arraystretch}{1.1}
  \setlength{\tabcolsep}{0pt}
  \setlength{\aboverulesep}{0pt}
  \setlength{\belowrulesep}{.15pt}
  \setlength{\extrarowheight}{0pt}
}

\definecolor{colOK}{HTML}{C5FFC2}
\definecolor{colNOTOK}{HTML}{EED9D8}

\tikzset{
every picture/.style={
	remember picture,
	baseline,
},
every node/.style={
	inner sep  = 0pt,
	anchor     = base,
	align      = center,
	text depth = .25ex,
  },
}

\newlist{RQs}{enumerate}{1}
\setlist[RQs]{
	topsep     = 1ex,
	parsep     = .2ex,
	itemsep    = .7ex,
	leftmargin = 2.2em,
	label      = {\textsmaller[1]{\textsf{RQ\textsubscript{\larger\arabic*}}}},
}
\Crefname{RQsi}{}{}

\newlist{objectives}{enumerate}{1}
\setlist[objectives]{
	topsep     = .7ex,
	parsep     = .2ex,
	itemsep    = 0pt,
	leftmargin = 2.6em,
	label      = {\textsmaller[1]{\bfseries(O\arabic*)}},
}
\Crefname{objectivesi}{Objective}{Objectives}

\newlist{assumptions}{enumerate}{1}
\setlist[assumptions]{
	topsep     = .5ex,
	parsep     = .1ex,
	itemsep    = 0pt,
	leftmargin = 3.3em,
	label      = \textsmaller[1]{\textsf{\bfseries A\arabic*~\:-}},
	ref        = \textscale{.9}{\textsf{A\arabic*}},
}
\Crefname{assumptionsi}{Assumption}{Assumptions}

\newlist{steps}{enumerate}{1}
\setlist[steps]{
	start      = 0,
	topsep     = .5ex,
	parsep     = .1ex,
	itemsep    = 0pt,
	leftmargin = 1.7em,
	label      = (\textscale{.93}{S}\arabic*),
	ref        = (\textscale{.93}{S}\arabic*),
}
\Crefname{stepsi}{Step}{Steps}

\let\emptyset\varnothing

\newcommand{\code}[2][]{\texttt{\ifthenelse{\equal{#1}{}}{}{\color{#1}}%
	\smaller[.5]#2}\xspace}
\newcommand{\bfcode}[2][]{\code[#1]{\bfseries #2}}
\newcommand{\acronym}[1]{\ensuremath{\textsc{\larger{#1}}}\xspace}

\newcommand{\Cpp}{C\raisebox{1pt}{\smaller\kern-.4pt+\kern-.8pt+}\xspace}
\newcommand{\IDE}{\acronym{ide}}     

\newcommand{\FOSS}{\acronym{foss}}   
\newcommand{\NUM}{\texttt{\protect{\#}}}

\newcommand{\EVM}{\acronym{evm}}  
\newcommand{\ETH}{\acronym{eth}}  
\newcommand{\SSC}{\acronym{ssc}}  
\newcommand{\attacker}{\ensuremath{\textsf{\itshape\smaller[.5]A}}\xspace}
\newcommand{\victim}{\ensuremath{\textsf{\itshape\smaller[.5]V}}\xspace}
\newcommand{\TP}{\acronym{tp}}  
\newcommand{\FP}{\acronym{fp}}
\newcommand{\TN}{\acronym{tn}}
\newcommand{\FN}{\acronym{fn}}
\newcommand{\GT}{\acronym{gt}}
\newcommand{\URV}{\acronym{urv}}  
\newcommand{\REE}{\acronym{ree}}  
\newcommand{\TD}{\acronym{td}}    

\newcommand{\Detecti}{\textsc{Detecti}\xspace}

\NewDocumentEnvironment{longcaption}{ O{.85\linewidth} }
  {\begin{minipage}{#1}\smaller}
  {\end{minipage}}

\definecolor{verylightgray}{rgb}{.97,.97,.97}

\lstdefinelanguage{Solidity}{
	keywords=[1]{anonymous, assembly, assert, balance, break, call, callcode, case, catch, class, constant, continue, constructor, contract, debugger, default, delegatecall, delete, do, else, emit, event, experimental, export, external, false, finally, for, function, gas, if, implements, import, in, indexed, instanceof, interface, internal, is, length, library, log0, log1, log2, log3, log4, memory, modifier, new, payable, pragma, private, protected, public, pure, push, require, return, returns, revert, selfdestruct, send, solidity, storage, struct, suicide, super, switch, then, this, throw, transfer, true, try, typeof, using, value, view, while, with, addmod, ecrecover, keccak256, mulmod, ripemd160, sha256, sha3}, 
	keywordstyle=[1]\color{blue}\bfseries,
	keywords=[2]{address, bool, byte, bytes, bytes1, bytes2, bytes3, bytes4, bytes5, bytes6, bytes7, bytes8, bytes9, bytes10, bytes11, bytes12, bytes13, bytes14, bytes15, bytes16, bytes17, bytes18, bytes19, bytes20, bytes21, bytes22, bytes23, bytes24, bytes25, bytes26, bytes27, bytes28, bytes29, bytes30, bytes31, bytes32, enum, int, int8, int16, int24, int32, int40, int48, int56, int64, int72, int80, int88, int96, int104, int112, int120, int128, int136, int144, int152, int160, int168, int176, int184, int192, int200, int208, int216, int224, int232, int240, int248, int256, mapping, string, uint, uint8, uint16, uint24, uint32, uint40, uint48, uint56, uint64, uint72, uint80, uint88, uint96, uint104, uint112, uint120, uint128, uint136, uint144, uint152, uint160, uint168, uint176, uint184, uint192, uint200, uint208, uint216, uint224, uint232, uint240, uint248, uint256, var, void, ether, finney, szabo, wei, days, hours, minutes, seconds, weeks, years},	
	keywordstyle=[2]\color{teal}\bfseries,
	keywords=[3]{block, blockhash, coinbase, difficulty, gaslimit, number, timestamp, msg, data, gas, sender, sig, value, now, tx, gasprice, origin},	
	keywordstyle=[3]\color{violet}\bfseries,
	identifierstyle=\color{black},
	sensitive=true,
	comment=[l]{//},
	morecomment=[s]{/*}{*/},
	commentstyle=\color{gray}\ttfamily,
	stringstyle=\color{red}\ttfamily,
	morestring=[b]',
	morestring=[b]"
}

\def\TITLESHORT{Vulnerability anti-patterns in Solidity}
\def\TITLELONG{Increasing smart contracts security by reducing false alarms}
\def\KEYWORDS{Software Security, Light-weight Code Analysis, Threats and Attack Modelling, Ethereum Smart Contracts, Usable Security and Privacy, Static Analysis for Security Vulnerabilities, Ethereum Smart Contracts, Security Vulnerability, Software Security Engineering, False Positives}
\hypersetup{%
    pdftitle={\TITLESHORT: \TITLELONG},
    pdfauthor={Tommaso Oss, Carlos E. Budde},
    pdfkeywords={\KEYWORDS},
}

\begin{document}

\title{\TITLESHORT: \TITLELONG}
\author{%
    Tommaso Oss\,\orcidID{0009-0005-2071-9551}\inst{1}
    \and
    Carlos E.\ Budde\,\orcidID{0000-0001-8807-1548}\inst{1}
}
\institute{University of Trento, Trento, Italy}

\authorrunning{Oss \& Budde}
\titlerunning{\TITLESHORT}


\maketitle

\setcounter{footnote}{1}  
\renewcommand{\thelstlisting}{\arabic{lstlisting}}  

\begin{abstract}
Turing completeness has made Ethereum smart contracts attractive to blockchain developers and attackers alike.
To increase code security, many tools can now spot most known vulnerabilities---at the cost of production efficiency.
Recent studies show false-positive ratios over 99\% in state-of-the-art technologies: this makes them impractical for use in industry and have raised questions on the direction of academic research.
In this work we show how integrating and extending current analyses is not only feasible, but also a next logical step in smart-contract security.
We propose light-weight static checks on the morphology and dynamics of Solidity code, stemming from a developer-centric notion of vulnerability, that we use to verify the output of other tools, flag potential false alarms, and suggest verifications.
Besides technical details we implemented an open-source prototype.
For three top-10 vulnerabilities it flags 324 warnings of other tools as false-positives, in 60 verified de-duplicated smart contracts selected from the blockchain by the presence of true (and false) vulnerabilities.
This amounts to a 92\%- to 100\%-reduction in the number of false-positives for these vulnerabilities.
\end{abstract}

\section{Introduction}
\label{sec:intro}

Since the practical solution of decentralised double-spending, the once academic question of cryptocurrencies' market viability now fuels one of the world's fastest developing technologies of the past 30 years \cite{Sza97}.
Today, cryptocurrencies are handled via digital ledgers known as \emph{blockchains}, which deploy transactions in distributed, unregulated, zero-trust ``crypto markets''.
Since its origins in 2008 \cite{Nak08}, blockchains market size rose over \$1\,B by 2022, and passed the \$1k\,B cap in 2024, with prognostics of equivalent growth for the foreseeable future \cite{CryptoMarket2}.

The first blockchain cryptocurrencies implemented cryp\-to\-graph\-ic\-ally-pro\-tec\-ted ledgers to register transactions among parties/contracts, where the latter exist off-chain.
Ethereum expanded this in 2014, turning \emph{smart contracts}---written in Turing-complete programming languages, mainly \emph{Solidity}---into first-class citizens of the blockchain.
Thus, besides account \textsc{\larger[.5]id}s and transaction amounts, this  blockchain encodes if-then-else clausal statements and loops for iterative payments \cite{But14,Woo14}.
This had two major consequences:
first, Ethereum's timeliness and versatility made it rise in rankings to the top-3 blockchains in both transactions-volume and market-cap, where it stands since 2015 \cite{CryptoMarket1,CryptoMarket2};
but also, \emph{security errors in Solidity smart contracts are public and unfixable}, waiting for any malicious actor to exploit them.  
This combination led to a constant stream of cyberattacks---worth millions of dollars \emph{each}---that continues to this day \cite{ATAAACK!}.

Response by the community resulted in a plethora of security-enhancing analyses and tools, ranging from static and dynamic checks of program code and bytecode, fuzz testing, and deep learning, to formal models of contracts interaction, and even game-theoretical analyses \cite{BLMZ22,WHZ+21,LLW+19}.
However and given the stakes, most of the effort has been on the identification of potential vulnerabilities.
More precisely, when designing a tool to determine whether certain portion of code is susceptible to malicious exploitation by a third party, \emph{true positives and detection sensitivity are the focus}.
This has led to a bloating of (false) alarms, where recent findings suggest false positive rates as high as 99.8\% for certain types of vulnerability \cite{ZZS+23}.
Thus, while the current state-of-the-art in Solidity smart contracts security can cover much of the vulnerability spectrum, \emph{its feasibility to integrate in production environments has yet to be proven}.

\begin{hlbox}[Research Question:]
\vspace{-.7ex}
	Can current methods and tools that check the security of
    Solidity smart contracts be used in modern code development?
\\[-1ex]
\end{hlbox}
\vspace{-1ex plus .5ex minus .1ex}

\noindent
We refine this research question into the following subquestions:
\vspace{-.5ex plus .3ex}
\begin{RQs}
\item	\label{RQ:tools_available}
		Are there freely available tools capable off-the-shelf of finding
		security vulnerabilities in Ethereum smart contracts?
\item	\label{RQ:tools_quality}
		What are the precision and specificity of these tools, to detect
        common security vulnerabilities in Solidity?
\item	\label{RQ:tools_enhancement}
		Can the precision (and specificity) of these tools
        be enhanced without compromising their ease of use?
\end{RQs}

\paragraph{Security vulnerabilities in smart contracts.}
\label{def:vulnerability}
We follow the standard terminology from \acronym{ict} practitioners in which \emph{security incidents} are set apart from safety incidents, setting our research focus in the former.
These involve an interplay between \emph{attackers} (\attacker) and \emph{victims} (\victim\!) in detriment of the latter, via vulnerabilities exploits and where $\attacker\neq\victim$ \cite{GI22}.
We focus on vulnerabilities detectable in-chain, e.g.\ exploitable by \attacker from existent code and regardless of actions by \victim.
While the canonical example in our setting is an illegitimate transfer of funds from \victim to \attacker, actions such as blocking functions in a contract that keep \victim from claiming rightful funds are also covered \cite{Yu.2021}. 
In contrast, attacks such as phishing or social engineering to gain illegitimate access to otherwise valid functionality (are undetectable from code inspection and) fall out of scope.
We henceforth refer to the above as our \emph{definition of security vulnerability} and highlight its lack of formal rigour, which we discuss further in \Cref{sec:bkg:related_work}.

We answer subquestions \Crefrange{RQ:tools_available}{RQ:tools_enhancement} for security vulnerabilities in Solidity smart contracts (\SSC), via a structured procedure yielding \emph{four contributions}:
%
%
\begin{enumerate}[leftmargin=1.2em,parsep=1ex,topsep=1ex]
\item	We first gather \SSC with different security vulnerabilities: 
        \begin{itemize}[parsep=.3ex,topsep=0pt,itemsep=0pt]
		\item	we guide our choice of vulnerabilities by the
				\acronym{owasp} top-10 \cite{OWASP};
		\item	we revise the literature for \SSC vulnerable to the above,
				with verified source code retrievable via official sources
				such as \mbox{\href{https://etherscan.io}%
				{\texttt{\color{blue}{etherscan.io}}}};
		\item	we iterate the two steps above, as we find that not all
				vulnerabilities have verified, de-duplicated \SSC with
				findable source code---\emph{we thus generate
				a dataset of 60 verified Solidity smart contracts
				with available source code, that the literature deems
				vulnerable to unchecked return value, reentrancy, and
				timestamp dependence (20 contracts per vulnerability)}%
				---see \Cref{sec:contracts};
		\end{itemize}
\item 	For \Cref{RQ:tools_available} we revise and select tools that detect vulnerabilities in \SSC:
        \begin{itemize}[parsep=.3ex,topsep=0pt,itemsep=0pt]
		\item	we use a two-level filter, first selecting tools
				based on four usability criteria that range from availability to output quality---\emph{we find and categorise 13 (out of 39) tools as eligible ``off-the-shelf'' to different degrees of usability};
		\item	we then exercised the eligible tools in our test environment,
                selecting those applicable to our contracts dataset,
                finally choosing 3 to experiment with.
        \end{itemize}
\item   We execute these tools in our dataset of 60 distinct \SSC, to determine the confusion matrix of each tool on for each vulnerability and answer \Cref{RQ:tools_quality}:
        \begin{itemize}[parsep=.3ex,topsep=0pt,itemsep=0pt]
        \item   our dataset is divided into three sets of 20 contracts,
				each set targeting a specific vulnerability and labelled
				with true- and false-positives;
		\item	\emph{for each set we run all tools, generate the corresponding
				confusion matrix, and compute precision and specificity},
				also measuring runtime.
        \end{itemize}
\item   We study the false positives above, identifying patterns incorrectly flagged by more than one tool to interpret them as ``anti-patterns'' recognisable via syntactic-checks, and thus propose an answer to \Cref{RQ:tools_enhancement}:
        \begin{itemize}[parsep=.3ex,topsep=0pt,itemsep=0pt]
		\item   \emph{We characterise patterns in Solidity source code that resemble three types of vulnerabilities but are in fact false-positives};
		\item	We implement a prototypical \FOSS tool (GPLv3, \cite{OB24}) to recognise these in code, and \emph{experiment in our dataset measuring its performance in precision, specificity, and runtime, showing the feasibility of our approach};
        \end{itemize}
\end{enumerate}

\paragraph{Data and Artifact Availability.}
We provide public access to all data and programs created and used in this work at \acronym{doi} 
\href{https://doi.org/10.6084/m9.figshare.26121655}{\textscale{.9}{\texttt{\color{blue}10.6084/m9.figshare.26121655}}}~\cite{OB24}.

\paragraph{Outline}
\Cref{sec:bkg} discusses related works after covering minimal background on smart-contract security, including the concrete vulnerabilities under study.
Then, \Cref{sec:contracts} describes our (necessarily manual) search for 60 \SSC vulnerable to the above, and \Cref{sec:tools} reviews \FOSS tools that we assessed for the experimentation done to answer \Cref{RQ:tools_available,RQ:tools_quality}.
Reverse-engineering the outcomes of these tools revealed the vulnerability anti-patterns that we introduce in \Cref{sec:invuln} to answer \Cref{RQ:tools_enhancement}, and that we use for the confusion matrix results presented in \Cref{sec:experiments}.
Finally, after discussing limitations and extensions in \Cref{sec:discussion}, this work concludes in \Cref{sec:conclu}.

\section{Background}
\label{sec:bkg}

\subsection{Security of Solidity smart contracts}
\label{sec:bkg:solidity}

\paragraph{Blockchain.}
A blockchain is a distributed ledger of \emph{blocks} (e.g.\ of transactions), chained via cryptographic technologies that make each block unique based on the chain preceding it.
Adding a new block is done via zero-trust consensus algorithms executed by independent \emph{nodes}, thus addressing deficiencies of centralised systems such as scalability limitations and single points of failure \cite{DiPierro}.

\paragraph{Ethereum.}
To form a chain, each block has a metadata header that minimally includes a timestamp and a hash of the previous block.
Depending on the implementation, the payload of a block (transaction) can have different types.
Besides transactions of its cryptocurrency (\ETH), the \emph{Ethereum} blockchain allows the payload to contain executable code.
Therefore, this code called \emph{smart contract} can be executed by any node in the blockchain~\cite{Zou}---and once the result of its execution enters the blockchain it becomes very hard or impossible to revert.  

\begin{wrapfigure}[7]{R}{.57\linewidth}
\vspace{-3.7ex}
\captionsetup[lstlisting]{justification=raggedleft,}
\begin{lstlisting}[%
    style=solidity,
	firstnumber=9,
    label={code:solidity_snippet},
    caption={Solidity snippet from \protect\href{https://etherscan.io/address/0x0744a686c17480b457a4fbb743195bf2815ca2b8\#code}{\ttfamily\color{blue}\underline{Eas}y\underline{Invest10}}} %, the full contract is in \cpageref{code:EasyInvest10}},
    ]
 owner.send(msg.value/5);`\label{code:snippet:TP}`
 if (invested[msg.sender] != 0){  
   address kashout = msg.sender;
   uint256 getout = invested[msg.sender]*10/100 *(block.number-atBlock[msg.sender])/5900;
   kashout.send(getout);`\label{code:snippet:FP}`
 }
\end{lstlisting}
\end{wrapfigure}

\paragraph{Solidity.}
Code execution of Ethereum smart contracts follows the semantics of the \emph{Ethereum Virtual Machine} (\EVM \cite{Woo14}).
But the executed bytecode was compiled from the programming language in which the contract was first written by humans.
From all languages available, \emph{Solidity} is officially recommended by Ethereum and, by far, the most popular for smart contract development \cite{OHJ20}.
\Cref{code:solidity_snippet} is a snippet containing Solidity builtin datatypes like \lstinline|address| (the address of a contract or wallet in the Ethereum blockchain), logical branching with \lstinline|if|, and the builtin function \lstinline|send| (to transfer funds between addresses)---see \Cref{app:solidity} for more examples.
\textquote{Influenced by \Cpp, Python, and JavaScript}, Solidity is Turing-complete and vulnerable to security bugs~\cite{Solidity}.

\paragraph{Vulnerabilities.}
Solidity smart contracts (\SSC) can be attacked by malicious parties, who exploit security vulnerabilities like in any stateful programming language \cite{PSS+21}.
Vulnerabilities are classified, e.g.\ Denial of Service or integer overflow, and statistics show the predominance of certain types both by the frequency with which they are found, as well as the likelihood of being exploited \cite{OWASP}.
The Open Worldwide Application Security Project (\acronym{owasp}) compiles this information: we study the following three vulnerability types, taken from the \acronym{owasp} top 10, which match our security focus---see \cite{OWASP} or \Cref{app:vulnerabilities} for a complete list.

\begin{itemize}[leftmargin=1em,topsep=.5ex,parsep=.7ex,itemsep=0pt]
\item	\textbf{Unchecked Return Value (\URV):}
%
Solidity functions \lstinline|call|, \lstinline|send|, \lstinline|callcode|, and \lstinline|delegatecall|\ return a Boolean stating the success of the operation, and then contract execution continues.
Regardless of the reason for a failure---e.g.\ stack overflow or lack of gas%
\footnote{%
\textquote[WHZ+21]{Ethereum requires users to pay for each step of the deployed contracts. The basic unit of the fee is called gas. [\,\ldots] initiators need to pay some gas for the execution}
}%
---attackers may exploit contracts whose logic ignores the return value of these functions and just continue execution.
\Cref{code:snippet:TP} in \Cref{code:solidity_snippet} is one such example, where the execution of the \lstinline|if| statement is oblivious of the return value of the \lstinline|send|, that it assumes successful.
This type of vulnerability is also known as Unchecked External Call.
%
\item	\textbf{Reentrancy (\REE):}
Reentrancy attacks leverage an insecure execution order of instructions in a victim contract, that allows an attacker to repeatedly call a function before it completes execution.
The classic example is an invocation to \lstinline|call|, to transfer \ETH to the user (attacker), before updating its balance in the contract.
This has been used to siphon all \ETH from a victim contract few transactions, making reentrancy the most feared Solidity vulnerability \cite{Yu.2021,ZZS+23}.
\item	\textbf{Timestamp Dependence (\TD):}
Some \SSC use the block timestamp (constants \lstinline|block.timestamp| or \lstinline|now|) to implement time-dependent logic such as locking funds for certain duration or seeding randomness for fairness. 
But the Ethereum protocol allows the timestamp of a block to be affected---in at bounded but flexible time range---by the node executing it \cite{Woo14,YTZ22}.
This makes the timestamp manipulable, which an attacker can exploit e.g.\ when it is used to control the transfer of cryptocurrency \cite{YTZ22}. 
%
\end{itemize}

\subsection{Related work}
\label{sec:bkg:related_work}

\SSC security has been a hot topic for a decade.
Best-practices to write secure code are available for developers, from sources such as the Ethereum's Solidity manual and the \acronym{owasp} vulnerabilities list \cite{Solidity,OWASP}.
From a theoretical viewpoint, game theory has been used to find Stackelberg- or Nash-equilibria conditions based on (nodes) mining incentives and blockchain consensus protocols \cite{LLW+19}.
Instead, we revise the state-of-the-art for \emph{vulnerabilities detection from \SSC code}~\cite{WHZ+21}.

\paragraph{\texorpdfstring{\SSC}{SSC} vulnerabilities.}
Regarding \emph{definitions of security vulnerability}, we find human-readable lists or taxonomies such as \acronym{owasp} and Chen et al.~(2021)~\cite{CPNX21}, or works that provide source- or byte-code definitions to identify vulnerable code fragments in practice---most of which result in too many \mbox{\FP~\cite{Dias.2021,GP20,PL21,ZZS+23}}.
\emph{We find no formal definition} for models with well-defined semantics---Time Automata, \acronym{mdp}{s}, etc.---actionable with formal tools support for verification.

While our definition in \Cref{sec:intro} takes a step in that direction, our focus is to help developers produce code not exploitable by 3\textsuperscript{rd} parties.
This removes the contract owner from the attackers (\attacker in \cpageref{def:vulnerability}), so code exploitable by hardcoded \lstinline|address|es are deemed \FP.
Such ``exploits'' are akin to common (legitimate) safety functions like \lstinline|selfdestruct| that the \lstinline|owner| can use to withdraw funds from the contract address.
Still, as restated in \Cref{sec:conclu}, the authors believe that the blockchain community would benefit from a formal definition of security vulnerability, actionable with modern model checkers with check times $<1\:s$.

\paragraph{Vulnerability detection.}
Most of the scientific literature studies how to increase detection sensitivity of \SSC vulnerabilities from byte- or source-code.
In surveys \cite{CPNX21,WHZ+21} we also find eight works that are closer to our research questions, in that they seek to increase (or at least measure) detection specificity.
We discuss them in \Cref{tab:related_work} and highlight that \cite{ZZS+23,HLSL23} served as main inspiration to this work: we propose generalisations that extend several of those principles to more vulnerabilities, minding runtime efficiency as per \Cref{RQ:tools_enhancement}.

\begin{table}[t]
	\centering
	\caption{Related works that measure or reduce \FP ratio}
	\label{tab:related_work}
	\smallskip
	\smaller
	\colcolortablepreamble
\def\heading#1{\multicolumn{1}{c}{\!\!\bfseries\vphantom{\larger[3]{I}}#1}}
\def\subheading#1{%
	\rowcolor{shade2}\midrule\multicolumn{3}{|c|}{\textrm{#1}}\\\midrule}
\begin{tabular}{C{.07\linewidth}@{~~}L{.425\linewidth}@{~~}L{.44\linewidth}}
\toprule
	\heading{Work}
	& \heading{Points in common}
	& \heading{Differences}
\\[.2ex]
\subheading{
	Practical \FP code pattern recognition for reentrancy
}
	\parbox[b]{\linewidth}{\centering%
		\cite{Yu.2021}\\[1ex]
		\cite{ZZS+23}
	}
	& Study the limitations of state-of-the-art tools to find reentrancy, proposing \FP code patterns. \cite{Yu.2021} further introduces a symbolic-execution tool that mitigates reentrancy \FP via filters based on these patterns.
	& We cover two vulnerabilities besides reentrancy, for which we too propose code patterns that we successfully used to detect \FP. We further target low (check) runtimes for integration in production environments (\Cref{RQ:tools_enhancement}).
\\
\subheading{
	Sound symbolic execution with practical comparisons
}
	\cite{Chang.2018}
	& Analyses control-flow graphs, looking for paths involving monetary transactions that violate security properties. This is a sound approach to find vulnerabilities that reduces the number of \FP w.r.t.\ other tools (see \Cref{sec:tools}).
	& The use of symbolic execution (a) produces paths that the user has to check to re-interpret as code, and (b) has limited coverage e.g.\ for loops. We perform static checks that point directly at code lines in an \SSC.
\\
\subheading{
	Existent tools efficacy: flagged (vulnerable) vs.\ exploited code
}
	\parbox[b]{\linewidth}{\centering%
		\cite{Dias.2021}\\[1ex]
		\cite{HLSL23}\\[1ex]
		\cite{PL21}
	}
	& Study the efficacy of existent tools to find \SSC vulnerabilities: they report high \FP ratios and a mismatch between flagged code and exploited code (the latter being as low as 1\% of the former for some vuln.). \cite[Fig.~5]{HLSL23} suggests \FP code patterns.
	& We are not only interested in studying other tools, but also in implementing the identified code patterns into a prototypical tool, to run experiments and measure the efficiency with which they can detect \FP.
\\
\subheading{
	Synthetic tool benchmark frameworks, including bug-injection
}
	\parbox[b]{\linewidth}{\centering%
		\cite{DFAC20}\\[1ex]
		\cite{GP20}
	}
	& Propose benchmarking approaches for \SSC security tools: the SmartBugs execution framework, and SolidiFI for bug-injection including several types of security vulnerabilities.
	& We do not target large synthetic datasets, but use other tools' results to identify \FP code patterns that we implement and test. We work on verified \SSC, deployed in Ethereum and labelled by humans as vulnerable.
\\
\bottomrule
\end{tabular}

\end{table}


\section{Finding vulnerable Solidity smart contracts}  
\label{sec:contracts}

\paragraph{Tools of the trade.}  
\Cref{sec:bkg} shows a well-known bloating in the number of code fragments (and contracts) flagged as vulnerable by state-of-the-art tools \cite{Dias.2021,HLSL23,ZZS+23}.
While this has been interpreted as evidence of a misalignment between academia and industry, the thesis of this work is that current technologies can be \emph{easily} and \emph{efficiently} adapted to reduce the number of false positives (\FP).

By \emph{easily} we mean that code inspection can reveal \FP patterns for each vulnerability type, that can be added to current and future software tools to reduce their \FP ratio.
This is linked to the definition of security vulnerability, which (can be tuned and) for us considers ``self-harm'' as \FP, since it is now known that it does not cause exploits---see \cref{sec:intro,sec:bkg:related_work}, and \cite{PL21,HLSL23}.

By \emph{efficiently} we mean that static checks---or other light-weight verification approach---should suffice to spot most \FP with minor runtime impact.
Ideally, these checks should be fast enough to allow developers to perform their regular tasks, while guiding their attention to code fragments that require a security-perspective revision.
As we show in \Cref{sec:experiments}, current state-of-the-art tools seem to go in either of these directions: speed or else precision.
By reducing the number of \FP with fast runtime we aim for faster precision.

\paragraph{A curated dataset of labelled \SSC source code.}
Two consequences of the above are:
(a) from the works that list thousands of vulnerable \SSC, a very low percentage contains contracts actually vulnerable; and
(b) to distinguish true vulnerable patterns from false ones---e.g.\ by human inspection---a dataset of \SSC with true security vulnerabilities must be built.
Also (c) most tools work with bytecode---see \Cref{sec:tools}---so building a varied dataset of \SSC \emph{source code examples truly affected by specific vulnerabilities} is non-trivial.
This means that practical studies like ours, that target the end developer, cannot easily bootstrap from existent datasets.
Our research requires \SSC deployed in the Ethereum blockchain---to operate on \emph{real \TP} as opposed to the typically easier-to-detect \emph{injected \TP}---with de-duplicated verified source code in \mbox{\href{https://etherscan.io}{\texttt{\color{blue}{etherscan.io}}}}, and (ideally human-verified) labels of security vulnerabilities.

\begin{wraptable}[8]{R}{.36\linewidth}
	\vspace{-4.5ex}
	\caption{Curated dataset}
	\label{tab:dataset}
	\centering\smaller\vspace{-2ex}
\begin{tabular}{lccc}
	\toprule
	Vulnerability      & \!\!\NUM{cntr.} & \TP &      \FP \\
	\midrule
	Unchk.\:Ret.\:Val. &       20     &      5 &      23 \\
	Reentrancy         &       20     &      3 &      79 \\
	Timestamp Dep.     &       20     &      5 &      67 \\
	\midrule
	\bfseries Total    &   \bf 60     & \bf 13 & \bf 169\\
	\bottomrule
\end{tabular}

\end{wraptable}

Therefore, we performed a systematic search on existent datasets for \SSC matching this criteria.
Per vulnerability in scope we selected 20 contracts, with verified source code available in the blockchain, among which at least three contain a vulnerability matching our definition.
To reduce bias, \TP and \FP labelling was performed for ten contracts per vulnerability by one of the authors, and verified by the other.
Specifically for \URV and \TD all contracts were taken from \cite{ScrawlD}---which contains automatic-tool labels only---while for \REE we also resorted to \cite{ZZS+23}---which has human labels whose \TP match with ours.

\begin{hlbox}
\vspace{-.5ex}
	\Cref{tab:dataset} summarises our dataset,
	which we make publicly available in \cite{OB24}.
\\[-3ex]
\end{hlbox}
\vspace{-1ex plus .5ex minus .1ex}

\section{FOSS tools for Solidity security}
\label{sec:tools}

\begin{table}
	\centering
    \caption{Criteria for selecting tools as eligible for use ``off-the-shelf''}
	\label{tab:tools-criteria}
	\smallskip
	\smaller[2]
	\colcolortablepreamble
\def\ok{\cellcolor{colOK!50}\,%
  \tikz{\node [circle,fill=colOK,draw=Green,thick,xshift=5pt,yshift=1pt]
    {\textscale{.8}{\checkmark}};}\xspace
}
\def\KO{\cellcolor{colNOTOK!50}\,%
  \tikz{\node [circle,fill=colNOTOK,draw=Red,thick,xshift=5pt,yshift=1pt]
    {\textscale{.8}{$\boldsymbol{\times}$}};}\xspace
}
\begin{tabular}{c@{~~}				
	>{\!\!}P{.2\linewidth}@{~~}		
	>{\!\!}P{.2\linewidth}@{~~}		
	>{\!\!}P{.22\linewidth}@{~~}	
	>{\!\!}P{.21\linewidth}}		
\toprule
	\vphantom{$\big)$}                        Level       
	& \multicolumn{1}{>{\!\!\!}c}{\bfseries C1: availability}
	& \multicolumn{1}{>{\!\!\!}c}{\bfseries C2: installation}
	& \multicolumn{1}{>{\!\!\!}c}{\bfseries C3: usage input}
	& \multicolumn{1}{>{\!\!\!}c}{\bfseries C4: usage output}
\\\midrule
	\textsf{\bfseries 1}
	& \ok Tool publicly available, with download link for unrestricted use
	& \ok Simple setup ($\leqslant5$ commands) via provided instructions
	& \ok Usage commands provided and well explained, succeeds on first try
	& \ok Output simple and clear, understandable at first glimpse
\\
	\textsf{\bfseries 2}
	& \KO Tool proposed, but not public or only upon contact with authors
	& \ok Complex setup ($>5$ commands) via provided instructions
	& \ok Usage commands provided but details missing, takes some trial and error
	& \ok Output too verbose or complex, but results of analysis are findable
\\
	\textsf{\bfseries 3}
	& \KO No tool: only a theoretical approach or algorithm is proposed
	& \ok Many setup processes proposed, only some work
	& \KO Usage commands not provided (or only some examples, hard to generalise)
	& \KO Output shows errors, hard or impossible to obtain results of analysis
\\
	\textsf{\bfseries 4/5}
	&
	& \KO (\textsf{\bfseries4}) Setup only possible via external web search, or (\textsf{\bfseries5}) not possible
\\\bottomrule
\end{tabular}

\end{table}


To answer \Cref{RQ:tools_available} we reviewed popular tools from online resources and surveys \cite{Mythril,WHZ+21}, to select those expected to be accessible to the standard developer.
We deem a tool \emph{usable off-the-shelf} if it is publicly available and easy to setup, execute, and understand its output.
We make our assessment systematic in \Cref{tab:tools-criteria}, which defines four criteria divided in levels:
a green cell marked
	\tikz{\node [circle,fill=colOK,draw=Green,thick,xshift=5pt,yshift=1pt]
	{\textscale{.7}{\checkmark}};}
indicates a criterion level eligible as off-the-shelf;
red cells marked
	\tikz{\node [circle,fill=colNOTOK,draw=Red,thick,xshift=5pt,yshift=1pt]
	{\textscale{.7}{$\boldsymbol{\times}$}};}
indicate the opposite%
\footnote{%
This is inspired in the artifact badging system of \acronym{cs} conferences, where e.g.\ the \acronym{acm} \emph{Available} badge is equivalent to our availability lvl.\,1, the \emph{Functional} badge requires all criteria to be at most lvl.\,2, and for the \emph{Reusable} badge all levels must be 1.
}.
Thus, eligible tools must have an availability level $\leqslant1$, installation level $\leqslant3$, and usage-input and usage-output levels $\leqslant 2$.


\begin{table}[t]
	\centering
    \caption{Evaluation of tools usable ``off-the-shelf''
	         by the criteria from \Cref{tab:tools-criteria}}
	\label{tab:tools-eligible} 
	\smallskip
	\smaller[2]
	\colcolortablepreamble
	\begin{minipage}[c][.34\textheight][t]{.51\linewidth}
\def\YEP{\checkmark}
\def\OK#1{\cellcolor{colOK}#1}
\def\BAD#1{\cellcolor{colNOTOK}#1}
\def\HELLYES{\cellcolor[HTML]{9AFF99}Yes}
\def\HELLNO{\cellcolor[HTML]{FFCCC1}No}
\begin{tabular}{l   
                C{2.1em}C{2.1em}C{2.1em}          
				C{1.5em}C{1.5em}C{1.5em}C{1.5em}  
				c}  
\toprule
& \multicolumn{3}{c}{Vuln.\;detect.} & \multicolumn{4}{c}{Criteria}  &
\\\cmidrule(l{2pt}r{2pt}){2-4}\cmidrule(l{2pt}r{3pt}){5-8}
\multirow{-2}{*}{\bfseries Tool}
               & \URV & \REE & \TD  &   C1    &   C2    &   C3    &   C4    & \multirow{-2}{*}{\bfseries Elig.}
\\\midrule
AChecker       &      &      &      & \OK{1}  & \OK{1}  & \OK{1}  & \OK{1}  & \HELLYES \\
ConFuzzius     & \YEP & \YEP & \YEP & \OK{1}  & \OK{3}  & \OK{2}  & \OK{2}  & \HELLYES \\
ContractFuzzer & \YEP & \YEP & \YEP & \OK{1}  & \BAD{5} & \BAD{-} & \BAD{-} & \HELLNO  \\
ContractWard   & \YEP & \YEP & \YEP & \BAD{2} & \BAD{-} & \BAD{-} & \BAD{-} & \HELLNO  \\
EasyFlow       &      &      &      & \OK{1}  & \BAD{5} & \BAD{-} & \BAD{-} & \HELLNO  \\
Echidna        &      &      &      & \OK{1}  & \OK{1}  & \OK{1}  & \OK{1}  & \HELLYES \\
EtherSolve     &      & \YEP &      & \OK{1}  & \OK{1}  & \OK{1}  & \OK{2}  & \HELLYES \\
Ethlint        & \YEP &      & \YEP & \OK{1}  & \OK{1}  & \OK{2}  & \OK{1}  & \HELLYES \\
eThor          &      & \YEP &      & \OK{1}  & \OK{1}  & \OK{2}  & \BAD{3} & \HELLNO  \\
ExGen          & \YEP &      &      & \OK{1}  & \BAD{5} & \BAD{-} & \BAD{-} & \HELLNO  \\
Gasper         &      &      &      & \BAD{2} & \BAD{-} & \BAD{-} & \BAD{-} & \HELLNO  \\
Halmos         &\tiny?&\tiny?&\tiny?& \OK{1}  & \OK{1}  & \BAD{3} & \BAD{-} & \HELLNO  \\
Harvey         & \YEP & \YEP &      & \OK{1}  & \BAD{5} & \BAD{-} & \BAD{-} & \HELLNO  \\
Horstify       & \YEP & \YEP & \YEP & \OK{1}  & \BAD{5} & \BAD{-} & \BAD{-} & \HELLNO  \\
MadMax         &      &      &      & \OK{1}  & \BAD{5} & \BAD{-} & \BAD{-} & \HELLNO  \\
Maian          &      &      &      & \OK{1}  & \BAD{5} & \BAD{-} & \BAD{-} & \HELLNO  \\
Manticore      &\tiny?&\tiny?&\tiny?& \OK{1}  & \BAD{5} & \BAD{-} & \BAD{-} & \HELLNO  \\
Medusa         &      &      &      & \OK{1}  & \OK{1}  & \OK{2}  & \OK{1}  & \HELLYES \\
Mythril        & \YEP & \YEP & \YEP & \OK{1}  & \OK{3}  & \OK{2}  & \OK{1}  & \HELLYES \\
\bottomrule
\end{tabular}

	\end{minipage}
	\begin{minipage}[c][.34\textheight][t]{.46\linewidth}
\def\YEP{\checkmark}
\def\OK#1{\cellcolor{colOK}#1}
\def\BAD#1{\cellcolor{colNOTOK}#1}
\def\HELLYES{\cellcolor[HTML]{9AFF99}Yes}
\def\HELLNO{\cellcolor[HTML]{FFCCC1}No}
\begin{tabular}{l   
                C{2.1em}C{2.1em}C{2.1em}          
				C{1.5em}C{1.5em}C{1.5em}C{1.5em}  
				c}  
\toprule
& \multicolumn{3}{c}{Vuln.\;detect.} & \multicolumn{4}{c}{Criteria}  &
\\\cmidrule(l{2pt}r{2pt}){2-4}\cmidrule(l{2pt}r{3pt}){5-8}
\multirow{-2}{*}{\bfseries Tool}
               & \URV & \REE & \TD  &   C1    &   C2    &   C3    &   C4    & \multirow{-2}{*}{\bfseries Elig.}
\\\midrule
Osiris         & \YEP & \YEP & \YEP & \OK{1}  & \OK{3}  & \OK{1}  & \OK{2}  & \HELLYES \\
Oyente         & \YEP & \YEP & \YEP & \OK{1}  & \OK{3}  & \OK{1}  & \OK{1}  & \HELLYES \\
ReGuard        & \YEP &      &      & \BAD{2} & \BAD{-} & \BAD{-} & \BAD{-} & \HELLNO  \\
Remix          & \YEP & \YEP & \YEP & \OK{1}  & \OK{1}  & \OK{1}  & \OK{1}  & \HELLYES \\
S-Gram         &\tiny?&\tiny?&\tiny?& \BAD{2} & \BAD{-} & \BAD{-} & \BAD{-} & \HELLNO  \\
Securify       & \YEP & \YEP & \YEP & \OK{1}  & \BAD{4} & \OK{1}  & \OK{1}  & \HELLNO  \\
Seraph         & \YEP &      &      & \BAD{2} & \BAD{-} & \BAD{-} & \BAD{-} & \HELLNO  \\
Sereum         & \YEP &      &      & \BAD{3} & \BAD{-} & \BAD{-} & \BAD{-} & \HELLNO  \\
sFuzz          & \YEP & \YEP & \YEP & \OK{1}  & \OK{2}  & \BAD{3} & \BAD{-} & \HELLNO  \\
Slither        & \YEP & \YEP & \YEP & \OK{1}  & \OK{1}  & \OK{1}  & \OK{2}  & \HELLYES \\
SmartCheck     & \YEP & \YEP & \YEP & \OK{1}  & \OK{1}  & \OK{1}  & \OK{1}  & \HELLYES \\
SmartCopy      & \YEP & \YEP & \YEP & \BAD{2} & \BAD{-} & \BAD{-} & \BAD{-} & \HELLNO  \\
SmartShield    & \YEP & \YEP &      & \BAD{2} & \BAD{-} & \BAD{-} & \BAD{-} & \HELLNO  \\
SoliDetector   & \YEP & \YEP & \YEP & \BAD{2} & \BAD{-} & \BAD{-} & \BAD{-} & \HELLNO  \\
Solscan        & \YEP & \YEP & \YEP & \OK{1}  & \OK{1}  & \OK{1}  & \OK{1}  & \HELLYES \\
Vandal         & \YEP & \YEP &      & \OK{1}  & \BAD{5} & \BAD{-} & \BAD{-} & \HELLNO  \\
VeriSmart      &\tiny?&\tiny?&\tiny?& \OK{1}  & \BAD{5} & \BAD{-} & \BAD{-} & \HELLNO  \\
Vultron        & \YEP & \YEP &      & \OK{1}  & \BAD{5} & \BAD{-} & \BAD{-} & \HELLNO  \\
WANA           & \YEP & \YEP & \YEP & \OK{1}  & \OK{1}  & \OK{1}  & \BAD{3} & \HELLNO  \\
Zeus           & \YEP & \YEP &      & \OK{1}  & \BAD{5} & \BAD{-} & \BAD{-} & \HELLNO  \\
\bottomrule
\end{tabular}

	\end{minipage}
\end{table}

Applying this criteria to the 39 tools found in the literature produces the results shown in \Cref{tab:tools-eligible}, which resulted in 13 eligible tools.
From those we further filter out the tools that cannot process some of the vulnerabilities targeted by our study, and that are unable to process more than 30\% of our curated dataset of \SSC (e.g.\ by incompatibility of the Solidity version)---full details on this second-level filter are in \Cref{app:eligible_tools_benchmark}.
We highlight that, albeit extensive, our search was not necessarily exhaustive---see \Cref{sec:conclu} for a discussion on this.

The final result were three tools that we used for experimentation in this work: Slither \cite{Slither}, Mythril \cite{Mythril}, and Remix \cite{Remix}.
Remix is the official \IDE for Solidity, developed and maintained by Ethereum.
While the other tools were specifically designed for security checking, Remix provides a much broader development functionality such as autocompletion and linting.
Our use of Remix is however limited to security checks, for which we considered both the compilation warnings displayed on the editor's gutter, as well as the alerts generated by the \emph{Solidity Analyzers} plugin.
In turn, this plugin can integrate three tools for code analysis: Remix Analysis (which covers the security category and we select), Solhint (a Solidity linter), and Slither (which we study independently).

\begin{hlbox}
\vspace{-.2ex}
	\Cref{RQ:tools_available} has a positive answer:
	\Cref{tab:tools-eligible} characterises 13 tools usable off-the-shelf.
\\[-3ex]
\end{hlbox}
\vspace{-1ex plus .5ex minus .1ex}

\section{Vulnerability anti-patterns}
\label{sec:invuln}

The vulnerabilities from \Cref{sec:bkg:solidity} can, in part, be automatically detected in \SSC using security tools like Slither, Mythril, and Remix.
This \namecref{sec:invuln} proposes simple code patterns which, when matched by a ``vulnerability'', suggests that it is in fact a \FP.
These results provide a (theoretical) positive answer to \Cref{RQ:tools_enhancement}, and were matured mainly by reverse-engineering the outcomes of our experiments---the rationale and some code examples are given in \Cref{app:invuln}.
Still, we introduce these patterns (first) here to ease the interpretation of results in \Cref{sec:experiments,sec:discussion}.

\subsection{Unchecked Return Value}
\label{sec:invuln:URV}

Studying the \FP of the tools tested we find five vulnerability ``anti-patterns'' that can be deemed false alarms.
Matching any of these patterns indicates a \FP:
\begin{enumerate}[leftmargin=1.2em,label=\textbf{\arabic*.},topsep=.5ex,parsep=.7ex]
\item If the visibility of the function is \lstinline|private| or \lstinline|internal|, and is unreachable from \lstinline|public| or \lstinline|external| functions free of restrictions (e.g.\ modifiers), they are inaccessible to attackers.
\item Similar to the above, if the potentially vulnerable function has modifiers that restrict its access to e.g.\ only the contract \lstinline|owner|, it is inaccessible to attackers.
\item If the flagged function contains restrictions that make it reachable only through the validation of special conditions (typically \lstinline|require|), user access is restricted; if it matches a hardcoded \lstinline|address| then it is inaccessible to attackers.
\item If the recipient address is the caller (viz.\ the address on which the call to transfer cryptocurrency is made), a failure can only result in self-harm: failed transfer cryptocurrency to themselves via \lstinline|call| or \lstinline|send|, or self-defined function via \lstinline|delegatecall| or \lstinline|callcode|. By our definition of vulnerability this is a \FP.
\item If the function is the last one in the function, no harmful subsequent instructions can be executed. Matches of \URV in this case are a \FP.
\end{enumerate}
\smallskip

\subsection{Reentrancy}
\label{sec:invuln:REE}

Experiments carried out in our curated dataset, in conjunction with the insights given in \cite{ZZS+23}, gave rise to the following six \REE anti-patterns:

\begin{enumerate}[leftmargin=1.5em,label=\textbf{\arabic*.},topsep=.5ex,parsep=.7ex]
\item[\bfseries1--3.\!\!\!\!\!]~~%
These patterns are the same as those listed for \URV in \Cref{sec:invuln:URV}, which for \REE also match the so-called ``Permission Control Group'' from \cite{ZZS+23}.
%
\addtocounter{enumi}{3}
\item If the address at which the external call is initialised is a hardcoded \lstinline|address|, it is beyond the control of the attacker. This means that a reentrancy exploit cannot be launched.
\item If no state variable is modified after the external call (e.g.\ the balance of a user) and before function termination, then reentrancy is not possible.
\item If the \lstinline|msg.value| to transfer is less than or equal to the amount of cryptocurrency received by the vulnerable function, reentrancy would at best increase rather than decrease funds at the victim contract, until the attacker runs out of funds---this matches the ``Special Transfer Value'' from \cite{ZZS+23}.
\end{enumerate}

\subsection{Timestamp Dependence}
\label{sec:invuln:TD}

This vulnerability differs from the other two in that the attacker is not a regular \SSC user calling the contract, but rather a miner that executes the contract in a transaction to be added to a block.
Notwithstanding, code analysis can still effectively find vulnerable contracts, and it can also yield false positives.
In particular we found one vulnerability anti-patter for \TD:

\begin{enumerate}[leftmargin=1.5em,label=\textbf{\arabic*.},topsep=.5ex,parsep=.7ex]
\item If the \lstinline|timestamp| is used (directly or via variables) \emph{exclusively} to verify presence in a time interval greater than 20 seconds, i.e.\ by a value comparison to constants or other variables, then by the ``15-seconds rule'' \cite{YTZ22} the likelihood of an attack is extremely low if not altogether impossible.
\end{enumerate}

%


\section{Empirical experiments on FP reduction}
\label{sec:experiments}

This \namecref{sec:experiments} demonstrates our pattern-recognition approach, via practical experiments on our curated dataset of 60 \SSC from the Ethereum blockchain.

As control we run them via the three selected tools (Slither, Mythril, and Remix) to find \URV, \REE, and \TD vulnerabilities, creating confusion matrices and measuring execution runtimes.
As test we implemented a prototypical tool, originally designed to refine the vulnerabilities found by these tools, but that here we execute standalone.
All our results and scripts are in a public artifact accessible at \acronym{doi} 
\href{https://doi.org/10.6084/m9.figshare.26121655}{\textscale{.9}{\texttt{\color{blue}10.6084/m9.figshare.26121655}}}~\cite{OB24}.

\subsection{The \texorpdfstring{\Detecti tool\hfill\href{https://github.com/Oss28/Detecti}{\texttt{\smaller\color{blue}https://github.com/Oss28/Detecti}}}{Detecti tool}}
\label{sec:experiments:Detecti}

We developed a Python tool, \href{https://github.com/Oss28/Detecti}{\Detecti (GPLv3)}, that runs static checks to identify the patterns presented in \Crefrange{sec:invuln:URV}{sec:invuln:TD}.
It uses Surya \cite{Surya} to parse the code of an input \SSC and identify potentially offending lines related to \URV, \REE, or \TD.
\Detecti can also function as semi-automatic wizard, offering the user the option to perform the distinct checks mentioned in \Cref{sec:invuln}.

Technically, for \URV it initially considers all instructions with the \lstinline|call|, \lstinline|send|, \lstinline|delegatecall|, or \lstinline|callcode| functions and an unused return value, deeming them \TP if none of the patterns from \Cref{sec:invuln:URV} are matched.
An analogous approach is followed for \REE, evaluating functions containing \lstinline|call| or instructions directly executing functions from other contracts.
For \TD it considers the instructions that can (directly or via variables) use the block timestamp as indicated.


\subsection{Results from smart contracts verifications}
\label{sec:experiments:results}

\Crefrange{fig:experiments:URV}{fig:experiments:TD} show the outcomes of execution of the tools in our dataset.
There is one chart per vulnerability, and each is divided into true- and false-positives (above the dotted line), and true- and false-negatives (resp.\ below).
The false predictions of a tool are in darker colours.
The absolute value of a bar indicates the case count.
For instance, the Ground Truth (\GT, taken from our dataset) for \URV in \Cref{fig:experiments:URV} reports 5 \TP and 23 \TN.
Note that in \Cref{sec:contracts} we called the latter \FP; here we call them \TN to highlight that they are code lines of interest for the respective vulnerability, that cannot be exploited.

\begin{figure}
	\vspace{-3ex}
	\centering
	\begin{subfigure}{.3\linewidth}
    	\centering
    	\caption{\URV}
    	\label{fig:experiments:URV}
    	\includegraphics[height=.18\textheight]{chart_URV}
	\end{subfigure}
	\begin{subfigure}{.3\linewidth}
    	\centering
    	\caption{\REE}
    	\label{fig:experiments:REE}
    	\includegraphics[height=.18\textheight]{chart_REE}
	\end{subfigure}
	\begin{subfigure}{.315\linewidth}
    	\centering
    	\caption{\TD}
    	\label{fig:experiments:TD}
    	\includegraphics[height=.18\textheight]{chart_TD}
	\end{subfigure}
	\caption{Confusion matrices for vulnerabilities detected in our dataset}
	\label{fig:experiments}
	\vspace{-2ex}
\end{figure}

The bars corresponding to the tools Slither through \Detecti count one \FP for each line flagged in the contract that is a \TN according to the \GT, and conversely for \FN and \TP.
So for instance Mythril in \Cref{fig:experiments:URV} indicates 14 \FP (lines marked as vulnerable that are \TN in the \GT), 5 \TP (vulnerable lines that coincide with the \GT), 1 \FN (lines marked as safe that are \TP in the \GT), and 9 \TN (safe lines that coincide with the \GT).
\Cref{app:more_results} shows individual results.

Cases missed by a tool, while essentially negatives and counted as such in \Cref{fig:experiments}, are not lines of code analysed but rather blind spots.
They indicate runtime errors (for Slither, Mythril, and Remix), or lines that we are aware that the pattern-recognition implemented in \Detecti is currently not catching.

\begin{wraptable}[7]{R}{.355\linewidth}
	\vspace{-4ex}
	\centering
	\caption{Runtimes (s)}
	\label{tab:runtimes}
	\vspace{-1ex}
	\smaller
	\begin{tabular}{lccr}
		\toprule
		\bfseries Vuln. & Slither & \Detecti & Mythril \\
		\midrule
		\bfseries \URV  &  11.5   & 12.3     &  16658 \\
		\bfseries \REE  &  12.1   & 16.1     &  35779 \\
		\bfseries \TD   &  12.5   & 13.8     &  55655 \\
		\bottomrule
	\end{tabular}
\end{wraptable}
Finally, the numbers above each tool name are the aggregated runtime in seconds that it took to execute it in the 20 contracts of a vulnerability in a standard laptop.
These were measured with the \code{time} \acronym{unix} command---we show them enlarged in \Cref{tab:runtimes}---except for Remix.
As described in \Cref{sec:tools}, Remix vulnerability detection cannot be done in batch: being an \IDE we must use its \acronym{gui} for each contract separately.
Therefore the 400~s reported are just an estimation, that count 20 seconds per contract to load it in the \IDE and launch the corresponding analysis tasks.

\paragraph{\texorpdfstring{\URV}{URV} results.}
\Cref{fig:experiments:URV} shows identical signatures of Slither and Remix for Un\-checked Return Value.
These tools deem vulnerable every instruction where the values of the relevant functions are not captured by a variable or used in a guard.
In the 20 \URV contracts of our dataset this heuristic matches all (5) \TP but also 21 \FP, resulting in precision $\sfrac{5}{26}=0.19$, and specificity $\sfrac{2}{23}\approx0.09$.
Mythril did better, possibly thanks to its taint-analysis capabilities, with precision $\sfrac{4}{18}=0.22$, and specificity $\sfrac{9}{23}\approx0.39$ with 9 \TN identified.
In contrast, by using the patterns from \Cref{sec:invuln:URV} \Detecti is able to identify all the \FP as such, maxing out both precision and specificity.
However, the \TP of \code{SmartBlockchainPro} was incorrectly assessed as a \FP, which results in a lower sensitivity.

\paragraph{\texorpdfstring{\REE}{REE} results.}
\Cref{fig:experiments:REE} tells a somewhat similar story for Reentrancy, where Slither and Remix flag as vulnerable every external call function listed in \Cref{sec:invuln:REE} for which a ``change of state'' follows through.
However, these tools consider events emission as a state change, contrary to the Solidity specification \cite{Solidity}.
As a result, Slither had 52 \FP but no \FN, and Remix 34 \FP and 2 \FN (this omits four contracts where Remix failed with an error, amounting to 21 missed cases), yielding precisions resp.\ $\sfrac{3}{55}\approx0.05$ and $\sfrac{1}{35}\approx0.03$, and specificities $\sfrac{27}{79}\approx0.34$ and $\sfrac{24}{58}\approx0.41$.
Mythril showed better performance, producing 48 \FP and 1 \FN (precision $\sfrac{2}{30}\approx0.07$ and specificity $\sfrac{48}{76}\approx0.63$) and failing to analyze only the \code{Dice} contract.
In contrast, using the patterns from \Cref{sec:invuln:REE} \Detecti confirmed as negative every \TN analysed, maxing out specificity.
Precision however was null, as \Detecti currently only consider calls within a single function and it missed (i.e.\ did not analyse the source code) of the three \TP.
This limitation---unrelated to code pattern recognition---is discussed in \Cref{sec:discussion:REE}.

\paragraph{\texorpdfstring{\TD}{TD} results.}
For Timestamp Dependence the approach of Remix is simple: every use of \lstinline|block.timestamp| is vulnerable.
This results in precision and specificity of $\sfrac{5}{72}\approx0.07$ and $0$ resp.
Instead, Slither flags ``only'' 60 \FP and correctly identifies 7 cases as \TN, for a precision and specificity of $\sfrac{5}{65}\approx0.08$ and $\sfrac{7}{67}\approx0.10$ resp.
As before, Mythril comes the better tool with 44 \FP but also 4 \FN, for a precision and specificity of $\sfrac{23}{57}\approx0.40$ and $\sfrac{1}{45}\approx0.02$ resp.
Also as before, \Detecti improves on all the previous values with one \FP and another \FN, for a precision and specificity of $\sfrac{4}{5}\approx0.80$ and $\sfrac{61}{62}\approx0.98$ resp. 
These results include the same technical limitation mentioned for \REE: for this vulnerability, \Detecti did not inspect inter-function calls that was relevant for \TD, omitting the analysis of code that contained 5 instructions including four \FP and one \TP.

\begin{wraptable}[9]{r}{.46\linewidth}
	\vspace{-3.5ex}
	\centering
	\caption{Precision and specificity of the tools in our experiments} 
	\label{tab:precision_and_specificity}
	\vspace{-1ex}
	\smaller
	\colcolortablepreamble
	\begin{tabular}{l@{~}
				>{~}c>{~}c
				>{~\columncolor{shade4}}c>{~\columncolor{shade4}}c
				>{~}c>{~}c
				>{~\columncolor{shade4}}c>{~\columncolor{shade4}}c}
	\toprule
	& \multicolumn{2}{>{~}c@{~}}{Slither}
	& \multicolumn{2}{>{~\columncolor{shade4}}c}{Mythril\,}
	& \multicolumn{2}{>{~}c@{~}}{Remix}
	& \multicolumn{2}{>{~\columncolor{shade4}}c}{\Detecti}
	\\[-.5ex]
	\multirow{-1}{*}{\bfseries Vuln.}
	& Pr & Sp  
	& Pr & Sp  
	& Pr & Sp  
	& Pr & Sp  
	\\
	\midrule                                               
	\bfseries \URV
	& .19 & .09  
	& .22 & .39  
	& .19 & .09  
	&  1  &  1   
	\\
	\bfseries \REE
	& .05 & .34  
	& .07 & .63  
	& .03 & .41  
	&  0  &  1   
	\\
	\bfseries \TD
	& .08 & .10  
	& .40 & .02  
	& .07 &  0   
	& .80 & .98  
	\\
	\bottomrule
\end{tabular}

\end{wraptable}
\Cref{tab:precision_and_specificity} summarises these results, showing a relatively poor performance of the selected tools---in terms of precision and specificity--to analyse our curated dataset for security vulnerabilities.
This was already apparent by looking at \Crefrange{fig:experiments:URV}{fig:experiments:TD}, and noting the high amounts of \FP for all vulnerabilities, coinciding with the findings of the latest literature also for the cases of the \URV and \TD vulnerabilities.

\begin{hlbox}
	\Cref{RQ:tools_quality} is answered---with a low performance of
	available tools---in \Cref{tab:precision_and_specificity}.
\end{hlbox}

For our final research question we first observe the performance in terms of runtime of existent tools, as shown in \Cref{fig:experiments,tab:runtimes}.
The time required to run a security analysis in an already loaded contract, by both Slither and Remix, was always less than 5 seconds.
This suggest that these tools could very likely be used in a production environment.
On the other side of the spectrum we have Mythril, which consistently overcome  the other tools in all cases of precision and specificity, but also required an average of 40 minutes to analyse each contract, with a minimum of 3 minutes and a maximum of about 3.5 hours.

The approach we took of identifying vulnerability-specific code patterns, mapping them to semantic use, exploits light-weight syntactic verifications in a non-general environment.
Since our checks (implemented in \Detecti) are on static source code, the achieved speed is comparable to that of Slither in all cases.
At the same time, for these vulnerabilities and in our dataset, the quality of the vulnerability-detection outcomes shows substantial improvements even in comparison to Mythril, and despite some technical and semantic limitations that we discuss in depth in \Cref{sec:discussion}.

\begin{hlbox}
	\Cref{RQ:tools_enhancement} has a positive answer, demonstrated by
	\Detecti in \Cref{fig:experiments,tab:precision_and_specificity}.
\end{hlbox}




\section{Discussion}
\label{sec:discussion}

\subsection{General performance of the tools}
\label{sec:discussion:analysis}

While Mythril was the tool to produce most \FN, it was also the best performing tool in terms of (low) \FP rate and overall specificity.
This is likely attributable to its dynamic-analysis approach, which result in a correct but incomplete analysis of the possible execution paths in the contract.
Unchecked vulnerable paths cannot be flagged, yielding false negatives.

Oppositely, the static analysers Remix and Slither can only check syntax, and take the conservative approach of flagging many suspicious instructions as vulnerable, leaving the final decision to the programmer.
This unavoidably produces a very high number of false positives, i.e.\ 92\% of the positives produced by these tools in our experiments are actually false positives.
Our conjecture is that so many (false) alarms can weary the developer, who would then disregard most warnings and having a negative net effect in smart contract security.

We highlight how such high \FP rate affects Remix, commended for its ability to detect true positives, but without equivalent treatment of precision indices \cite{ZZS+23}.
Our experiments reveal that between Slither and Remix, the former has a (slightly) better general performance as it generates a few less false alarms.

More importantly, our results with \Detecti suggest that static analyses can indeed be enhanced to remediate a non-trivial amount of \FP, without compromising the responsiveness of the tool (as it happened e.g.\ with Mythril).
It could discern 95\% false positives out of the positive results given by the other tools.
Thus, as a proof of concept, \Detecti performed well to improve the specificity of existent tools: we study each vulnerability type next.

\subsection{Threats to validity and technical limitations of \texorpdfstring{\Detecti}{Detecti}}
\label{sec:discussion:limitations}

\paragraph{General technical limitations of the prototype.}
We examine the limitations stemming from the tests conducted in \Cref{sec:experiments}, as well as the technical implementation choices of the \Detecti prototype.
\begin{itemize}[leftmargin=1.2em,topsep=.5ex,parsep=.5ex]
\item The first limitation affects all vulnerabilities, and involves the tool's inability to display the names of complex variables in a humanly understandable format (see our artifact \cite{OB24}). In such instances, the output contains a placeholder instead of the variable name, with the option to print the variable's dictionary identification by activating the verbose option.
%
\item Another technical limitation is when default functions of Solidity such as \lstinline|send,call,transfer| are overloaded in the contract code.
The tool will not detect these alterations and might inaccurately assess uses as \TP.
For example for \URV, the current implementation of \Detecti assumes that \lstinline|send|, \lstinline|call|, \lstinline|callcode|, and \lstinline|delegatecall| are not overloaded.
While we have not found a single \SSC where these functions had been overloaded, if they are and e.g.\ the overloaded \lstinline|send| raises an exception instead of returning a \lstinline|false| upon failure, then a \URV vulnerability cannot occur for this function.
%
\item Also and as mentioned in \Cref{sec:experiments:results}, the current implementation of \Detecti performs analyses local to a function scope.
This was the cause of the missed vulnerabilities for \REE and \TD in \Cref{fig:experiments:REE,fig:experiments:TD}.
While the solution is apparent, namely creating and traversing the whole control flow graph, this would most likely result in long checking times that defeat a main driver of this research.
We are currently exploring partial solutions, that include a bounded number of functions hops which could cover most typical cases.
\end{itemize}

\paragraph{Improvements for \URV analysis.}
Identification of false-positives for this vulnerability type could be further improved as follows
\begin{itemize}[leftmargin=1.2em,topsep=.5ex,parsep=.5ex]
\item Check whether the analyzed code is in the \SSC constructor: this would label it as \FP because the constructor is executed only at contract deployment, and its runtime bytecode will not appear in the blockchain.
%
\item Refining the criteria for the verification conditions and modifiers. The current prototype assumes that conditions and modifiers invariably restrict access. While this is the case in all contracts found so far, it cannot be guaranteed (just as in the case of function overloading discussed above) without further code analysis. We expect that simple conditions such as keyword matches could be cheap and mostly effective---for a sound analysis the only solution may be \acronym{smt} solving \`a la Mythril.
%
\item As mentioned above, the function-scope search of \Detecti will fail if the vulnerability is distributed among functions. The \FN example is trivial: a function $f_A$ which contains a \URV but has \lstinline|private| visibility. If $f_A$ is called from another---\lstinline|public|---function $f_B$, and state modification occurs in $f_B$, then the contract should be flagged as vulnerable.
%
\end{itemize}


\paragraph{Improvements for \REE analysis.}
\FP identification could improve by:
\begin{itemize}[leftmargin=1.2em,topsep=.5ex,parsep=.5ex]
\item The same cases mentioned for \URV about \SSC constructor, inter-function calls, and conditions and modifiers. As a special case, ``cross-contract reentrancy'' is a variant of \REE that \Detecti cannot currently detect.
%
\item Builtin Solidity functions like \lstinline|send| and \lstinline|transfer| are usually deemed \FP for \REE due to the 2300 forwarded gas limit. It could be possible to overload e.g.\ \lstinline|send| such that when the fallback function is called it would forward all remaining gas, enabling a reentrancy attack. A mitigation to cover such cases could check that calls to builtin functions involve the expected parameters.
%
\end{itemize}


\paragraph{Improvements for \TD analysis.}
\FP identification could improve by:
\begin{itemize}[leftmargin=1.2em,topsep=.5ex,parsep=.5ex]
\item The same cases mentioned above about constructors and inter-function calls.
%
\item The current version of the tool does not go beyond the mechanisms to apply the ``15-second Rule''.%
\footnote{%
The 15-seconds rule is an approximative estimator of the exploitability of a time-based vulnerability, which states that \textquote[YTZ22]{If the scale of your time-dependent event can vary by 15 seconds and maintain integrity, it is safe to use a block.timestamp}
}
\Detecti identifies the scenario of time-interval verification, and heuristically attempts to establish whether the integrity of the operation of the contract cannot be compromised by the feasible manipulation of the timestamp.
While this is effective to spot \FP, a taint-analysis could extend the approach even to instructions that use the timestamp in operations related to values of cryptocurrency transfers.
\item On a similar note, even if a potential dangerous \TD is found, the peculiarities of this vulnerability make an impact analysis feasible and useful. For a simple example, fuzzy testing could be used to experiment with the magnitude of the manipulations of the timestamp on the amount of cryptocurrency transferred.
\end{itemize}

\section{Conclusions}
\label{sec:conclu}

This work extends the state-of-the-art in security revisions for Ethereum smart contracts written in Solidity.
In our literature search, we were able to find 13 \FOSS tools (including Slither, Mythril, and Remix) that can be used off-the-shelf for Solidity security enhancement, answering \Cref{RQ:tools_available} positively.

We exercised the three tools mentioned above on 60 smart contracts taken from the Ethereum blockchain, which we manually inspected and labelled to determine the presence of three types of vulnerabilities selected from the \acronym{owasp} top 10: Unchecked Return Value, Reentrancy, and Timestamp Dependency.

We exercised the tools to detect vulnerabilities in this dataset: comparing their diagnoses with our ground truth we found that 92\% of the vulnerabilities flagged by the tools are in fact false positives.
This gives a discouraging answer to \Cref{RQ:tools_quality}, as we conjecture that producing too-high a volume false positives can be more detrimental to overall company-efficiency and contract security (tired eyes will not pay attention to every ``useless'' warning).

From these results and via literature research, most notably \cite{WHZ+21,ZZS+23}, we proposed static analyses concerning our selected vulnerabilities, whose observation can increase the confidence with which specific lines or instructions are deemed potentially vulnerable, by flagging patterns related to false security vulnerability alarms.
This can improve practical usability of security tools via a lightweight pattern-recognition (static) approach.

The above gives a positive theoretical answer to \Cref{RQ:tools_enhancement}.
We further added an empirical demonstration by implementing a Python tool, \Detecti, and executing it on the same dataset as the other tools used for experimentation.
The outcomes were that \Detecti could detect most of the false positives generated by three state-of-the-art tools---including the official \IDE of Solidity, Remix.

These results take special relevance in the face or recent findings like \cite{ZZS+23}, which could have suggested a misdirection of academic endeavours regarding security enhancements of smart contracts.
We believe that our results, which leverage the efforts of much previous work, show how a proper integration of research in the field can provide much better practical performance for use in production environments.


\subsection*{Future work}
While we strived to cover most available \FOSS tools for Solidity security, the fast pace of the field makes it infeasible to guarantee complete coverage.
Many tools are deployed as a successful proof-of-concept whose maintenance stops after a few years (e.g.\ Oyente); others are currently under development or require non-trivial setups, essentially misaligning with our \Cref{RQ:tools_available}.
Still, we believe that a comprehensive experimental analysis of tools, under a well-defined but laxer selection criteria, would benefit the community.

In particular this would necessitate a catalogue or taxonomy of security vulnerabilities in smart contracts, against which the tools would be tested.
For the current work we resorted to well-known and official but ``informal'' resources such as the OWASP ranking \cite{OWASP}.
A hardened catalogue of vulnerabilities, e.g.\ with examples of contract snippets and descriptions of the exploits, is a \textit{sine qua non} for a more formal and comprehensive approach to the task.

Besides a taxonomy of vulnerabilities, a dataset of labelled vulnerable \SSC, real in the sense that they exist in the Ethereum blockchain, are fundamental for the generation of consistent benchmarks.
We contributed to this effort with our dataset, and are working to include further contracts in the future.

Moreover and as stated in \Cref{sec:intro,sec:bkg:related_work}, the community would greatly benefit from a formal definition of security vulnerability, ideally adaptable to the different variants of security interpretation e.g.\ as listed in works like \cite{CPNX21}.

Finally, regarding \Detecti, we have had the opportunity in previous sections to discuss how it would be possible to improve and complete the detection of patterns currently being considered. But it is worth mentioning that another strength of \Detecti is its modularity, which we hope will encourage other researchers to contribute and improve this tool, implementing new pattern detection, and considering new types of vulnerabilities.

\begin{credits}
\paragraph{Data availability.}
We provide public access to all data and programs created and used in this work at \acronym{doi} 
\href{https://doi.org/10.6084/m9.figshare.26121655}{\textscale{.9}{\texttt{\color{blue}10.6084/m9.figshare.26121655}}}~\cite{OB24}.
\paragraph{Funding.}
This work was funded by the European Union under
the \href{https://ec.europa.eu/info/research-and-innovation/funding/funding-opportunities/funding-programmes-and-open-calls/horizon-europe/marie-sklodowska-curie-actions_en}{MSCA} grant 101067199 (\textsmaller{\scshape ProSVED}),
and
\href{https://next-generation-eu.europa.eu/index_en}{NextGenerationEU} projects D53D23008400006 (\textsmaller{\scshape Smartitude}) under MUR PRIN 2022,
and
PE00000014 (\textsmaller{\scshape SERICS}) under MUR PNRR.
Views and opinions expressed are those of the author(s) only and do not necessarily reflect those of the European Union or The European Research Executive Agency.
Neither the European Union nor the granting authority can be held responsible for them.
\end{credits}

\bibliographystyle{splncs04}
\bibliography{paper} 

\begin{thebibliography}{10}
\providecommand{\url}[1]{\texttt{#1}}
\providecommand{\urlprefix}{URL }
\providecommand{\doi}[1]{https://doi.org/#1}

\bibitem{BLMZ22}
Bartoletti, M., Lande, S., Murgia, M., Zunino, R.: Verifying liquidity of
  recursive {Bitcoin} contracts. {Logical Methods in Computer Science}
  \textbf{18}(1) (2022). \doi{10.46298/lmcs-18(1:22)2022}

\bibitem{But14}
Buterin, V.: {Ethereum}: A next-generation smart contract and decentralized
  application platform (2014),
  \url{https://ethereum.org/content/whitepaper/whitepaper-pdf/Ethereum_Whitepaper_-_Buterin_2014.pdf}

\bibitem{Chang.2018}
Chang, J., Gao, B., Xiao, H., Sun, J., Cai, Y., Yang, Z.: sCompile: Critical
  Path Identification and Analysis for Smart Contracts, pp. 286--304 (10 2019).
  \doi{10.1007/978-3-030-32409-4_18}

\bibitem{CPNX21}
Chen, H., Pendleton, M., Njilla, L., Xu, S.: A survey on {Ethereum} systems
  security: Vulnerabilities, attacks, and defenses. {ACM} Comput. Surv.
  \textbf{53}(3),  67:1--67:43 (2021). \doi{10.1145/3391195}

\bibitem{CryptoMarket1}
Coinmarketcap (2024), \url{https://coinmarketcap.com/}

\bibitem{CryptoMarket2}
\texttt{crypto.com} (2024), \url{https://crypto.com/price}

\bibitem{ATAAACK!}
Hacks to cryptocurrencies in 2024 (2024),
  \url{https://www.immunebytes.com/blog/list-of-largest-crypto-hacks-in-2024/}

\bibitem{Medusa}
Crytic: Crytic/medusa: Parallelized, coverage-guided, mutational solidity smart
  contract fuzzing, powered by go-ethereum,
  \url{https://github.com/crytic/medusa}

\bibitem{DiPierro}
Di~Pierro, M.: What is the blockchain? Computing in Science \& Engineering
  \textbf{19}(5),  92--95 (2017). \doi{10.1109/MCSE.2017.3421554}

\bibitem{Dias.2021}
Dia, B., Ivaki, N., Laranjeiro, N.: An empirical evaluation of the
  effectiveness of smart contract verification tools. In: {PRDC}. pp. 17--26
  (2021). \doi{10.1109/PRDC53464.2021.00013}

\bibitem{Ethlint}
duaraghav8: Duaraghav8/ethlint: (formerly solium) code quality \& security
  linter for solidity, \url{https://github.com/duaraghav8/Ethlint}

\bibitem{DFAC20}
Durieux, T., Ferreira, J.F., Abreu, R., Cruz, P.: Empirical review of automated
  analysis tools on 47,587 {Ethereum} smart contracts. In: {ICSE}. pp.
  530--541. {ACM} (2020). \doi{10.1145/3377811.3380364}

\bibitem{Slither}
Feist, J., Grieco, G., Groce, A.: Slither: A static analysis framework for
  smart contracts. In: {WETSEB} (2019). \doi{10.1109/WETSEB.2019.00008}

\bibitem{GP20}
Ghaleb, A., Pattabiraman, K.: How effective are smart contract analysis tools?
  {E}valuating smart contract static analysis tools using bug injection. In:
  {ISSTA}. pp. 415--427. {ACM} (2020). \doi{10.1145/3395363.3397385}

\bibitem{AChecker}
Ghaleb, A., Rubin, J., Pattabiraman, K.: Achecker: Statically detecting smart
  contract access control vulnerabilities. In: {ICSE} (2023).
  \doi{10.1109/icse48619.2023.00087}

\bibitem{GI22}
Gibson, D., Igonor, A.: Managing Risk in Information Systems. Learning
  Information Systems Security \& Assurance, Jones \& Bartlett, 3 edn. (2022)

\bibitem{Surya}
GNSPS: Surya: A set of utilities for exploring solidity contracts (2019),
  \url{https://github.com/Consensys/surya}

\bibitem{Echidna}
Grieco, G., Song, W., Cygan, A., Feist, J., Groce, A.: Echidna: Effective,
  usable, and fast fuzzing for smart contracts. {SIGSOFT}  (2020).
  \doi{10.1145/3395363.3404366}

\bibitem{HLSL23}
Hu, T., Li, J., Storhaug, A., Li, B.: Why smart contracts reported as
  vulnerable were not exploited? TechRxiv  (2023).
  \doi{10.36227/techrxiv.21953189.v3},
  \url{https://www.techrxiv.org/doi/full/10.36227/techrxiv.21953189.v1}

\bibitem{LLW+19}
Liu, Z., Luong, N.C., Wang, W., Niyato, D., Wang, P., Liang, Y.C., Kim, D.I.: A
  survey on {Blockchain}: A game theoretical perspective. {IEEE} Access
  \textbf{7},  47615--47643 (2019). \doi{10.1109/ACCESS.2019.2909924}

\bibitem{Oyente}
Luu, L., Chu, D.H., Olickel, H., Saxena, P., Hobor, A.: Making smart contracts
  smarter (2016). \doi{10.1145/2976749.2978309}

\bibitem{Nak08}
Nakamoto, S.: Bitcoin: A peer-to-peer electronic cash system  (2008)

\bibitem{OHJ20}
Oliva, G.A., Hassan, A.E., Jiang, Z.M.J.: An exploratory study of smart
  contracts in the ethereum blockchain platform. Empir. Softw. Eng.
  \textbf{25}(3),  1864--1904 (2020). \doi{10.1007/S10664-019-09796-5}

\bibitem{OB24}
Oss, T., Budde, C.E.: Vulnerability anti-patterns in {Solidity}: {Detecti}
  artifact (experimental reproduction package) (2024).
  \doi{10.6084/m9.figshare.26121655},
  \url{https://figshare.com/articles/software/26121655}

\bibitem{OWASP}
Owasp smart contract top 10 (2023),
  \url{https://owasp.org/www-project-smart-contract-top-10/}

\bibitem{EtherSolve}
Pasqua, M., Benini, A., Contro, F., Crosara, M., Dalla~Preda, M., Ceccato, M.:
  Enhancing ethereum smart-contracts static analysis by computing a precise
  control-flow graph of ethereum bytecode. Journal of Systems and Software
  \textbf{200},  111653 (Jun 2023). \doi{10.1016/j.jss.2023.111653}

\bibitem{PL21}
Perez, D., Livshits, B.: Smart contract vulnerabilities: Vulnerable does not
  imply exploited. In: {USENIX Security 21}. pp. 1325--1341. {USENIX}
  Association (2021),
  \url{https://www.usenix.org/conference/usenixsecurity21/presentation/perez}

\bibitem{PSS+21}
Prana, G.A.A., Sharma, A., Shar, L.K., Foo, D., Santosa, A.E., Sharma, A., Lo,
  D.: Out of sight, out of mind? how vulnerable dependencies affect open-source
  projects. Empirical Software Engineering  \textbf{26}(4) (2021).
  \doi{10.1007/s10664-021-09959-3}

\bibitem{Solscan}
Riczardo: Riczardo/solscan: Static solidity smart contracts scanner written in
  python, \url{https://github.com/riczardo/solscan}

\bibitem{Mythril}
Sharma, N., Sharma, S.: A survey of {Mythril}, a smart contract security
  analysis tool for {EVM} bytecode. Indian Journal of Natural Sciences
  \textbf{13}(75) (2022)

\bibitem{Solidity}
Solidity programming language (2024), \url{https://docs.soliditylang.org/en}

\bibitem{ScrawlD}
Sujeetc: Sujeetc/scrawld, \url{https://github.com/sujeetc/ScrawlD}

\bibitem{Sza97}
Szabo, N.: Formalizing and securing relationships on public networks. First
  Monday  \textbf{2}(9) (1997). \doi{10.5210/fm.v2i9.548}

\bibitem{Remix}
Team, R.: Remix project: Jump into web3 (2022),
  \url{https://remix-project.org/}

\bibitem{SmartCheck}
Tikhomirov, S., Voskresenskaya, E., Ivanitskiy, I., Takhaviev, R., Marchenko,
  E., Alexandrov, Y.: Smartcheck: Static analysis of ethereum smart contracts.
  In: {WETSEB}. pp. 9--16. {ACM} (2018). \doi{10.1145/3194113.3194115}

\bibitem{ConFuzzius}
Torres, C.F., Iannillo, A.K., Gervais, A., State, R.: Confuzzius: A data
  dependency-aware hybrid fuzzer for smart contracts. In: EuroS\&amp;P (2021).
  \doi{10.1109/eurosp51992.2021.00018}

\bibitem{Osiris}
Torres, C.F., Sch\"{u}tte, J., State, R.: Osiris: Hunting for integer bugs in
  ethereum smart contracts. In: {ACSAC} (2018). \doi{10.1145/3274694.3274737}

\bibitem{WHZ+21}
Wang, Y., He, J., Zhu, N., Yi, Y., Zhang, Q., Song, H., Xue, R.: Security
  enhancement technologies for smart contracts in the blockchain: A survey.
  Trans Emerging Tel Tech  \textbf{32}(12), ~29 (2021).
  \doi{https://doi.org/10.1002/ett.4341}

\bibitem{Woo14}
Wood, G.: {Ethereum}: A secure decentralised generalised transaction ledger.
  Ethereum project yellow paper  \textbf{151},  1--32 (2014),
  \url{https://ethereum.github.io/yellowpaper/paper.pdf}

\bibitem{YTZ22}
Yaish, A., Tochner, S., Zohar, A.: Blockchain stretching {\&} squeezing:
  Manipulating time for your best interest. In: {EC}. pp. 65--88. {ACM} (2022).
  \doi{10.1145/3490486.3538250}

\bibitem{Yu.2021}
Yu, R., Shu, J., Yan, D., Jia, X.: {ReDetect}: Reentrancy vulnerability
  detection in smart contracts with high accuracy. In: {MSN}. pp. 412--419
  (2021). \doi{10.1109/MSN53354.2021.00069}

\bibitem{ZZS+23}
Zheng, Z., Zhang, N., Su, J., Zhong, Z., Ye, M., Chen, J.: Turn the {Rudder}: A
  beacon of reentrancy detection for smart contracts on {Ethereum}. In: {ICSE}.
  pp. 295--306. {IEEE} (2023). \doi{10.1109/icse48619.2023.00036}

\bibitem{Zou}
Zou, W., Lo, D., Kochhar, P.S., Le, X.B.D., Xia, X., Feng, Y., Chen, Z., Xu,
  B.: Smart contract development: Challenges and opportunities. {IEEE TSE}
  \textbf{47}(10),  2084--2106 (2021). \doi{10.1109/TSE.2019.2942301}

\end{thebibliography}

\appendix

\section{The Solidity language}
\label{app:solidity}

\begin{figure}
\centering
\begin{lstlisting}[%
  style=solidity,
  label={code:example},
  caption=Solidity code of an example contract,
  ]
// SPDX-License-Identifier: MIT
pragma solidity ^0.8.0;
contract CollectMoney {
    address payable public owner;
    constructor() {
        owner = payable(msg.sender);
    }
    modifier onlyOwner() {
        require(msg.sender == owner, "Not the owner");
        _;
    }
    function withdraw(uint amount) public onlyOwner {
       if(amount <= address(this).balance){
            _sendFunds(owner, amount);
        }
    }
    function _sendFunds(address payable recipient, uint amount) private {
        recipient.send(amount);
    }
    function getBalance() public view returns (uint) {
        return address(this).balance;
    }
}
\end{lstlisting}
\end{figure}

\Cref{code:example} shows the syntax of a Solidity smart contract.
Its structure resembles a class in object-oriented programming, where the main construct (a \lstinline|contract|) defines variables and functions and can be instantiated.
Local variables can also be defined internally to a \lstinline|function|.
Functions implement the interface by which external entities can interact with the contract (\lstinline|public| functions), as well as internal functionality (\lstinline|private| functions).

Solidity supports elementary datatypes such as Booleans, integers such as \lstinline|int8| and \lstinline|uint256|, and strings, as well as composite types such as arrays and \lstinline|struct|.
For control flow the usual branch and loop constructs are available (\lstinline|if|, \lstinline|while|), and recursive calls are possible.

The global namespace, and certain predefined types, expose variables and functions to read data from the blockchain, the \EVM global state, or to interact with other contracts/accounts.
Relevant examples include the functions \mbox{\lstinline|call|,} \lstinline|send|, and \lstinline|transfer|---mostly used to transfer \ETH cryptocurrency---and the variables \lstinline|block.number| and \lstinline|.timestamp|.

Moreover, a \lstinline|modifier| struct in Solidity can implement assertions (\lstinline|require|) such as the valid caller of a function; and a logging mechanism can be used to \lstinline|emit| events, i.e. notifications that are recorded on the blockchain. While they are useful for tracking the execution and behavior of the contract, they cannot by themselves modify the contract's state.


\section{Top-10 vulnerabilities in Solidity} 
\label{app:vulnerabilities}

The Open Worldwide Application Security Project (\acronym{owasp}) maintains a ranking of the ten most pervasive and exploited vulnerabilities in smart contracts \cite{OWASP}.
The list is broadly construed, including e.g.\ faulty code logic and mining attacks that are oblivious of the code in a smart contract.

From those vulnerabilities, \Cref{sec:bkg:solidity} selects three that can be (partially) identified via code patterns, thus falling in the scope of this work.
For completeness, here we present a brief overview of the entire top-10 ranking updated to 2023:

\begin{description}
   \item[1. Reentrancy Attacks:] A malicious contract exploits a loophole in the victim contract to repeatedly withdraw funds. Mainly caused by failing to promptly verify the exploiter's updated balance.
   \item[2. Integer Overflow and Underflow:] Each integer type has a range, so each integer variable can only store numbers within the range of its data types.  If this range is exceeded, it will cause an overflow or an underflow.
   \item[3. Timestamp Dependence:] Dependence on the block timestamp value to carry out an operation. Since the timestamp value is generated by the node executing the smart contract, it becomes susceptible to manipulation and vulnerable to attacks.
   \item[4. Access Control Vulnerabilities:] A type of security flaw that allows users without permission to interact with and alter data or functions in a smart contract.
   \item[5. Front-running Attacks:] The gas price of new transactions are appropriately adjusted for impacting the ordering of transactions waiting in the pool to be mined as blocks.
   \item[6. Denial of Service (DoS) Attacks:] A DoS attack involves exploiting vulnerabilities to exhaust resources such as gas, CPU cycles, or storage, rendering the contract unusable.
   \item[7. Logic Errors:] If a smart contract is poorly coded, it may contain logic errors that lead to unintended behavior. This could range from incorrect calculations to faulty conditional statements, or even exposed administrative functions.
   \item[8. Insecure Randomness:] Use of an insecure randomness source that allows attackers to manipulate or predict the generated number in order to gain benefits illegally.
   \item[9. Gas Limit Vulnerabilities:] When attackers exploit the gas limits to disrupt the normal functioning of the blockchain network. They can design transactions or smart contracts that consume an excessive amount of gas intentionally.
   \item[10. Unchecked External Calls:] The \code{call()}, \code{send()}, \code{callcode()}, and \code{delegatecall()} functions return a Boolean that states the success or failure of the operation. If the return value of the call is not checked, program execution could continue by assuming its success, even if call failed.
\end{description}

\section{Selection of eligible tools for this study}
\label{app:eligible_tools_benchmark}

After an initial filter of \SSC security analysis tools by the criteria defined in \Cref{sec:tools}---see \Cref{tab:tools-criteria,tab:tools-eligible}---a total of 13 tools were found eligible as usable ``off-the-shelf''.
We then assessed those tools in more depth, to select those that are applicable to the dataset of vulnerable \SSC presented in \Cref{sec:contracts}.

\Cref{tab:tools-benchmark} presents the results of this second evaluation, where three tools were found to be adequate for our needs in order to perform balanced and comprehensive tests, namely Slither, Mythril, and Remix Analysis.
In the case of rejected tools, the rightmost column describes the motivations that led to this choice.


\begin{table}
	\centering
	\caption{\FOSS tools for \SSC security that match our needs}
	\label{tab:tools-benchmark}
	\smaller[2]
	\def\ALIGN{\centering}
\def\TOOL#1{\parbox[t]{3em}{\ALIGN\protect{#1}\\[.5ex]\protect{\cite{#1}}}}
\def\NOPE{\parbox[t]{2em}{\ALIGN No}}
\def\YEP{\parbox[t]{2em}{\ALIGN\bfseries Yes}}
\begin{tabular}{>{\!\!}C{3em}@{~~}L{.34\linewidth}@{}c@{~~}L{.46\linewidth}}
\toprule
	\bfseries Tool
	& \bfseries Description
	& \bfseries \!\!\!\!Match
	& \bfseries Main reason
\\\midrule\arrayrulecolor{black!50}
\TOOL{Oyente}
	& Symbolic execution. Parses source- and byte-code.
	& \NOPE
	& The latest supported version is solc 0.4.19, which means really limited in terms of Solidity versions supported.
\\\midrule[.3pt]
\TOOL{Mythril}
	& Symbolic execution, \acronym{smt} solving, and taint analysis. Severity rating.\!
	& \YEP
	& Swift installation through Docker image, and vulnerabilities match our interest set. Covers all versions of Solidity (project maintained).
\\\midrule[.3pt]
\TOOL{Osiris}
	& Symbolic execution and taint analysis (leverages Oyente). 
	& \NOPE
	& Being based on Oyente it has the same problems as the latter.
\\\midrule[.3pt]
\TOOL{Slither}
	& Static checker and taint analysis, via the SlithIR intermediate representation.
	& \YEP
	& Super easy installation and use, and vulnerabilities math our interest set. Covers all versions of Solidity (project maintained).
\\\midrule[.3pt]
\parbox[t]{3em}{\ALIGN Smart\\ Check\\[.5ex]\cite{SmartCheck}}
	& Static checker via an XML-based intermediate representation.
	& \NOPE
	& Deprecated since 2020, failing for Solidity v0.6.0 and above, which reduces severely the test set of smart contracts.
\\\midrule[.3pt]
\TOOL{Remix}
	& Static checker (Remix Analysis is a plugin of Remix, the official \IDE of Solidity).
	& \YEP
	& Most functional out-of-the-box tool found, offered via package managers and even via an online version. Vulnerabilities match our interest set. Covers all versions of Solidity (project maintained).
\\\midrule[.3pt]
\TOOL{Echidna}
	&  Fuzzer to detect violations in assertions and custom properties.
	& \NOPE
	& Not exactly a vulnerability detector, it does not deal with any of the vulnerabilities of interest to us.
\\\midrule[.3pt]
\TOOL{ConFuzzius}
	&  Hybrid fuzzing, a combination of symbolic execution and fuzzing.
	& \NOPE
	& Complex command to launch analysis (many arguments). Returns errors for newer versions of Solidity.
\\\midrule[.3pt]
\TOOL{Solscan}
	&  Static checker based on regular expressions and contextual analysis.
	& \NOPE
	& For many contracts return errors similar to "An error occurred while checking NAME\_VULN. This vulnerability class was NOT checked." for some vulnerabilities (incompatible Solidity versions?).
\\\midrule[.3pt]
\TOOL{Ethlint}
	&  Static checker with a set of core rules for linting code.
	& \NOPE
	& Deprecated since 2019, failing for newer Solidity versions. Does not cover Reentrancy.
\\\midrule[.3pt]
\TOOL{Medusa}
	&  Go-ethereum-based fuzzer inspired by Echidna.
	& \NOPE
	& Being based on Oyente it has the same problems as the latter.
\\\midrule[.3pt]
\TOOL{AChecker}
	&  Static data-flow and symbolic-based analysis.
	& \NOPE
	& Focus on Access Control Vulnerabilities, Does not cover the three vulnerabilities of our interest.
\\\midrule[.3pt]
\TOOL{EtherSolve}
	&  Static checker, based on symbolic execution of the EVM operands stack.
	& \NOPE
	& It does not cover URV and TD. Analyzes only EVM bytecode, no Solidity source code.
\\\arrayrulecolor{black}\bottomrule
\end{tabular}

\end{table}

\section{Vulnerability anti-patterns by example}
\label{app:invuln}

\Cref{sec:invuln} introduced code patterns which, when flagged by a tool as a potential security vulnerability, suggest that it is in fact a false-positive.
Those patterns were discovered in part by reading the literature, but also by analysing the security analyses results of the tools used in our experimentation.
Here we detail these results, to give further insight into the reasons why the proposed patterns are indicative of \FP, with examples on Solidity smart contracts included in our curated dataset \cite{OB24}.

\subsection{Unchecked Return Value}
\label{app:invuln:URV}

Static checkers such as Slither and Remix flag all instances where the return value of an invocation to \lstinline|send|, \lstinline|call|, \lstinline|callcode|, or \lstinline|delegatecall| is not used in a guard or captured by a variable.
This simplistic syntactic match produces a significant amount of \FP, as an unchecked value does not imply an insecure contract logic, and this can be checked statically in many cases as we show next.

Note that experiments with Mythril exhibited better precision---although this depends strongly on the logic and size of the contract---possibly thanks to its taint-analysis capabilities.
The price to pay when using Mythtil was a much-longer runtime, several minutes per contract, in some cases even an hour or more.
This makes it hard to embed in production environments that implement e.g.\ agile code development.

\begin{figure}
\centering
\begin{lstlisting}[%
    style=solidity,
    label={code:Escrow},
    caption=Solidity smart contract \protect\href{https://etherscan.io/address/0x073e957bc883693f15ecb14bfced3e8ffc8654c5\#code}{\ttfamily\color{blue}\underline{Escrow}},
    ]
contract Escrow {
  address buyer;
  address seller;
  address arbitrator;
  function Escrow() payable {
    seller = 0x5ed8cee6b63b1c6afce3ad7c92f4fd7e1b8fad9f;
    buyer = msg.sender;
    arbitrator = 0xabad6ec946eff02b22e4050b3209da87380b3cbd;
  }
  function finalize() {
    if (msg.sender == buyer || msg.sender == arbitrator)
      seller.send(this.balance);`\label{code:Escrow:FP1}`
  }  
  function refund() {
    if (msg.sender == seller || msg.sender == arbitrator)
      buyer.send(this.balance);`\label{code:Escrow:FP2}`
  }
}
\end{lstlisting}
\vspace{-2ex}
\end{figure}

Regarding our \FP code patterns for the \URV vulnerability, \Cref{code:Escrow}  displays two examples.
More in detail, an Unchecked Return Value attack attempts to exploits the continuation of a contract execution that assumed (without checking) a successful end of instruction.
But in \cref{code:Escrow:FP1,code:Escrow:FP2} of \Cref{code:Escrow} nothing can continue executing, for the simple fact that there are no further instructions.
Note that, in spite of this, all tools mark the contract as vulnerable.

For another example we note that \cref{code:EasyInvest10:FP1} in \Cref{code:EasyInvest10} matches the third pattern discussed in \Cref{sec:invuln:URV}.
Because \code{kashout} is initialised as the \lstinline|msg.sender| address, which is the caller of the function.
Therefore, for attackers it makes no sense to have the crypto transfer to themselves fail if the following part of the contract will be executed regardless of the outcome of the transfer.
This is in close connection to our definition of security vulnerability, that requires that $\attacker\cap\victim=\emptyset$.

\begin{figure}
\centering
\begin{lstlisting}[%
    style=solidity,
    label={code:EasyInvest10},
    caption=Solidity smart contract \protect\href{https://etherscan.io/address/0x0744a686c17480b457a4fbb743195bf2815ca2b8\#code}{\ttfamily\color{blue}\underline{Eas}y\underline{Invest10}},
    ]
contract EasyInvest10 {
  address owner;
  constructor() {
    owner = msg.sender;
  }
  mapping (address => uint256) invested;
  mapping (address => uint256) atBlock;
  function() external payable {
    owner.send(msg.value/5);
    if (invested[msg.sender] != 0){  
      address kashout = msg.sender;
      uint256 getout = invested[msg.sender]*10/100*(block.number-atBlock[msg.sender])/5900;
      kashout.send(getout);`\label{code:EasyInvest10:FP1}`
    }
    atBlock[msg.sender] = block.number;
    invested[msg.sender] += msg.value;
  }
}
\end{lstlisting}
\vspace{-2ex}
\end{figure}

\subsection{Reentrancy}
\label{app:invuln:REE}

For reentrancy vulnerabilities, both Slither and Remix exhibit a similar operational approach, necessitating the presence of a specific condition for vulnerabilities of this type to be identified.
Our experiments match this with the modification of the contract state after an invocation, beyond the contract's scope, to \lstinline|call| or after a direct call to an external contract function, usually made by conversion of an address variable to an abstract contract type defined in the solidity file (\bfcode[blue]{abstr\_contr(adr).funct}).

There are two ways the state change can happen according to the criteria of these tools: either by emitting an event, or altering the value of some state variables.
However, the documentation specifies that events emission are logging primitives that cannot modify the state \cite{Solidity}, rendering all first cases moot.
In turn, and as pointed out in \cite{ZZS+23}, altering a state variable after such instructions is necessary but not sufficient to make the contract vulnerable to reentrancy attacks.
This provides another simple false positive check, which however cannot be directly used to spot true positives.

Take as an example \Cref{code:Decore}, where \cref{code:Decore:FP1} is a true state modification, precisely of the balance of a potential attacker, and it is executed after an invocation to \lstinline|call| two lines earlier.
Notwithstanding, this function is not vulnerable to reentrancy either, by virtue of the \code{onlyOwner} modifier, which asserts (via \lstinline|require|) that only the owner of the contract account is able to execute the function.

\begin{figure}
\centering
\begin{lstlisting}[%
    style=solidity,
    label={code:Decore},
    caption=Solidity code snippet of the \protect\href{https://etherscan.io/address/0xb944B46Bbd4ccca90c962EF225e2804E46691cCF\#code}{\ttfamily\color{blue}\underline{Decore}} contract,
    ]
contract Ownable is Context {
    $\ldots$
    modifier onlyOwner() {
        require(_owner == _msgSender(), "Ownable: caller is not the owner");
        _;
    }
}
$\ldots$
contract NBUNIERC20 is Context, INBUNIERC20, Ownable {
  $\ldots$
  function emergencyDrain24hAfterLiquidityGenerationEventIsDone() public onlyOwner {
    require(contractStartTimestamp.add(4 days) < block.timestamp, "Liquidity generation grace period still ongoing"); // About 24h after liquidity generation happens
    (bool success, ) = msg.sender.call.value(address(this).balance)("");
    require(success, "Transfer failed.");
    _balances[msg.sender] = _balances[address(this)];`\label{code:Decore:FP1}`
    _balances[address(this)] = 0;
  }
  $\ldots$
}
\end{lstlisting}
\vspace{-2ex}
\end{figure}

Finally we note that here, and as with unchecked return values, Mythril proved to be the most specific tool in our experimentation, at the cost of longer runtimes.
In contrast, our experiments with Remix and Slither resulted in 59\% and 66\% false positive rates.

\subsection{Timestamp Dependence}
\label{app:invuln:TD}

For \TD, Remix reports every use of \lstinline|block.timestamp| and/or \lstinline|now| as a potential vulnerability without apparent further analysis.
In contrast, the other tools seem to consider the use---direct or via variable declaration---of the timestamp as part of a Boolean condition in a branching, looping guard, or boolean variable definition.
Slither is conservative, marking any such occurrence as a vulnerable instruction.
Mythril performs stricter checks, and in fact our experiments in \Cref{sec:experiments:results} show a number of false negatives, where it fails to recognise a potentially vulnerable use of the timestamp.

For an example of the pattern introduced in \Cref{sec:invuln:TD} we draw attention to \cref{code:FifteenPlus:FP1} in \Cref{code:FifteenPlus}, where \code{prtime[owner]} is subtracted from \lstinline|now| (a syntax sugar for \lstinline|block.timestamp|), and then verified that the result is greater than or equal to 86400.
Therefore, it is checked that the timestamp of the block is at least one day later than the time instant (timestamp of a previous block where the contract was executed defining that variable) saved in \lstinline{prtime[owner]}.
A 15-seconds manipulation in this scenario would not be relevant, and thus flagging this use of \lstinline|now| as vulnerable is a \FP.

\begin{figure}
\begin{lstlisting}[%
    style=solidity,
    label={code:FifteenPlus},
    caption=Solidity code snippet of the \protect\href{https://etherscan.io/address/0xcfd2047eb61412e9b8de511dc2087e07003829ee\#code}{\ttfamily\color{blue}\underline{FifteenPlus}} contract,
    ]
contract FifteenPlus {
  $\ldots$
  mapping (address => uint256) timestamp;
  $\ldots$
  function() external payable {
    if((now-prtime[owner]) >= 86400){ `\label{code:FifteenPlus:FP1}`
            owner.transfer(ths.balance / 100);
            prtime[owner] = now;
    }
    $\ldots$
  }
}
\end{lstlisting}
\vspace{-2ex}
\end{figure}

We also highlight how \TD attacks are generally less severe than \REE or \URV.
Technically, value manipulations in this context are limited to few units/second, so even if the manipulable value (modulo arithmetic operations) is taken as a loss, that loss is bounded, in contrast to the other two vulnerabilities studied.
Nonetheless, our definition of vulnerability does include as \TP any manipulation of cryptocurrency that an attacker could exploit via \TD exploits.

\section{Detailed results per contract}
\label{app:more_results}

\begin{landscape}
\begin{figure}[t]
    \centering
	\begin{subfigure}{\linewidth}
		\centering
		\caption{Contracts potentially vulnerable to Unchecked Return Value}
		\label{fig:chart-results:urv}
		\includegraphics[width=\linewidth]{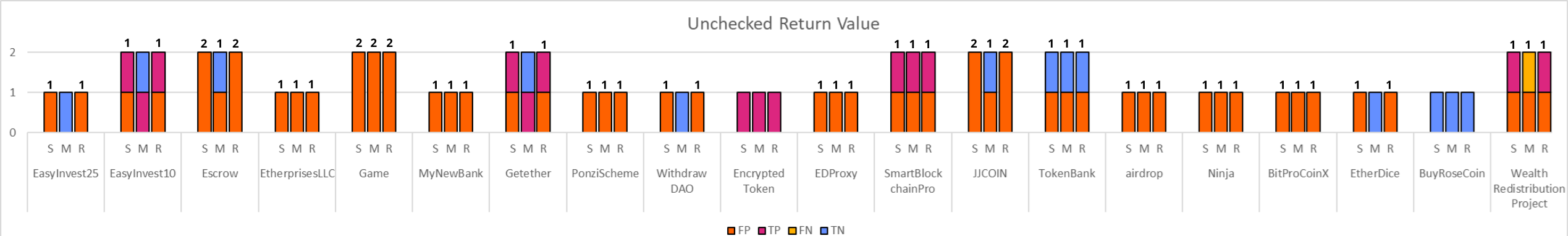}
	\end{subfigure}
	\begin{subfigure}{\linewidth}
		\centering
		\caption{Contracts potentially vulnerable to Reentrancy}
		\label{fig:chart-results:ree}
		\includegraphics[width=\linewidth]{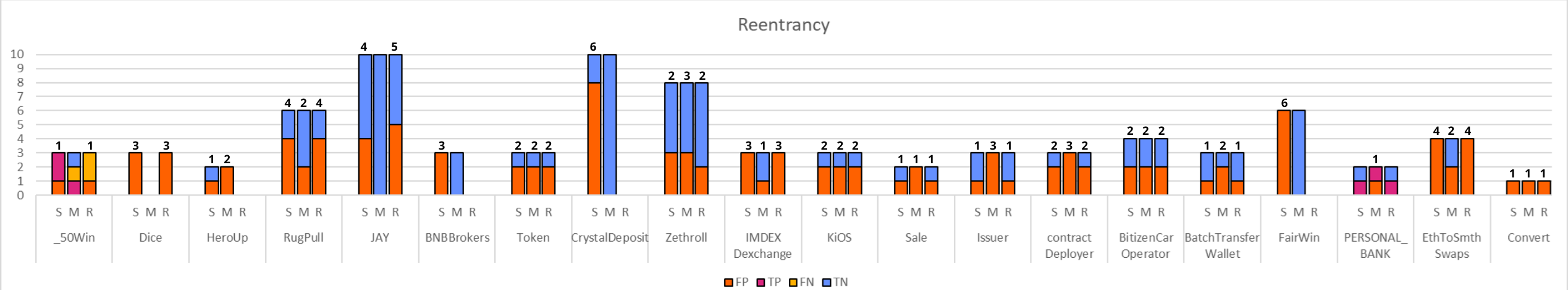}
	\end{subfigure}
	\begin{subfigure}{\linewidth}
		\centering
		\caption{Contracts potentially vulnerable to Timestamp Dependence}
		\label{fig:chart-results:td}
		\includegraphics[width=\linewidth]{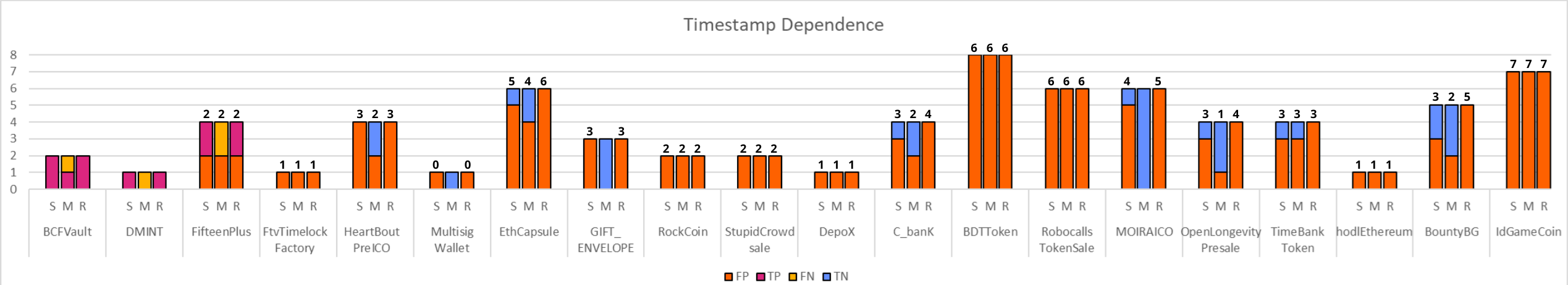}
	\end{subfigure}
    \caption{Individual results of our experiments, per contract}
    \label{fig:chart-results}
\end{figure}
\end{landscape}

\Cref{sec:experiments:results} presented results of our experiments, aggregated for all contracts in our dataset for each vulnerability.
Here, \Crefrange{fig:chart-results:urv}{fig:chart-results:td} show the outcomes for each individual contract.
The number on top of the columns represents the number of false positives, produced by the tool considered for the specific contract analyzed, that \Detecti correctly recognized as such.

\end{document}